\documentclass[twocolumn]{aastex631}

\usepackage{amsmath}

\begin{document}

\title{Blowing star formation away in AGN Hosts (BAH) - IV: Feeding and feedback in 3C\:293 observed with JWST NIRSpec}

\author[0009-0008-2184-1403]{Maitê S. Z. de Mellos}
\affiliation{Departamento de F\'isica, CCNE, Universidade Federal de Santa Maria, Av. Roraima 1000, 97105-900,  Santa Maria, RS, Brazil}

\author[0000-0003-0483-3723]{Rogemar A. Riffel}

\author[0009-0005-0583-5773]{Gabriel L. Souza-Oliveira}
\affiliation{Departamento de F\'isica, CCNE, Universidade Federal de Santa Maria, Av. Roraima 1000, 97105-900,  Santa Maria, RS, Brazil}
\affiliation{Centro de Astrobiología (CAB), CSIC-INTA, Ctra. de Ajalvir km 4, Torrejón de Ardoz, E-28850, Madrid, Spain}

\author[0000-0001-6100-6869]{Nadia L. Zakamska}
\affiliation{Department of Physics \& Astronomy, Johns Hopkins University, Bloomberg Center, 3400 N. Charles St, Baltimore, MD 21218, USA}

\author[0000-0002-6570-9446]{Marina Bianchin}
\affiliation{Department of Physics and Astronomy, 4129 Frederick Reines Hall, University of California, Irvine, CA 92697, USA}

\author[0000-0003-1772-0023]{Thaisa Storchi-Bergmann}

\author[0000-0002-1321-1320]{Rogério Riffel}
\affiliation{Departamento de Astronomia, IF, Universidade Federal do Rio Grande do Sul, CP 15051, 91501-970, Porto Alegre, RS, Brazil}

\author[0000-0003-3667-9716]{José Henrique Costa-Souza}
\affiliation{Departamento de F\'isica, CCNE, Universidade Federal de Santa Maria, Av. Roraima 1000, 97105-900,  Santa Maria, RS, Brazil}
\affiliation{Centro de Astrobiología (CAB), CSIC-INTA, Ctra. de Ajalvir km 4, Torrejón de Ardoz, E-28850, Madrid, Spain}

%\author[0000-0001-6100-6869]{Nadia L. Zakamska}
%\affiliation{Department of Physics \& Astronomy, Johns Hopkins University, Bloomberg Center, 3400 N. Charles St, Baltimore, MD 21218, USA}

%\author[0000-0003-1772-0023]{Thaisa Storchi-Bergmann}
%\author[0000-0002-1321-1320]{Rog\'erio Riffel}
%\affiliation{Departamento de Astronomia, IF, Universidade Federal do Rio Grande do Sul, CP 15051, 91501-970, Porto Alegre, RS, Brazil}

%\author[0000-0002-6570-9446]{Marina Bianchin}
%\affiliation{Department of Physics and Astronomy, 4129 Frederick Reines Hall, University of California, Irvine, CA 92697, USA}



\begin{abstract}
We use JWST/NIRSpec observations of the radio galaxy 3C\,293 to map the emission, extinction, and kinematics of hot molecular and ionized gas, as well as stellar kinematics, within the inner {$\sim 2 \rm \,kpc$}. The stellar velocity field is well described by a rotating disk model, with its kinematical center offset by $\sim 0.5 \rm \, arcsec$ from the continuum peak. The hot molecular gas is traced by the $\rm H_2\,2.12\,\mu m$ emission line, and the ionized gas by [Fe\,{\sc ii}]$\,\rm1.64 \mu m$ and $\rm Pa\alpha$. The gas presents three main kinematic components: a rotating disk seen as a narrow component ($\sigma \sim 100 \rm \, km \, s^{-1}$); a blueshifted broad outflow ($\sigma \sim 250$ km\,s$^{-1}$); and a fast ionized outflow as a very broad component ($\sigma\sim 640$ km\,s$^{-1}$). Extinction maps reveal high $A_V$ values, up to $\sim$\,35, spatially coincident with dust lanes seen in optical images. In addition to the disk and outflows components, inflows along the dust lanes are detected in H$_2$ gas, with a mass inflow rate of $\dot{M}_{\rm in}\simeq 4 \times 10^{-4} \rm ~M_{\odot} \, yr^{-1}$,  which is lower than the AGN accretion rate. For the outflows, we derive peak mass-outflow rates of $0.08 \pm 0.02 ~\rm M_{\odot}~yr^{-1}$ (molecular) and $6.5\pm 1.7 \rm ~M_{\odot} \, yr^{-1}$ (ionized). The outflow, driven by the radio jet, has a kinetic power of 5.7\% of the jet power -- enough to suppress star formation. Our results highlight 3C\,293’s turbulent post-merger history and JWST’s unique capability to probe dust-obscured AGN.

\end{abstract}



\section{Introduction} \label{sec:intro}

%Feedback de AGN, galaxias com excesso de emissao de H2, (trabalhos do Ogle, Petric, etc, incluindo na 3C\,293...
%resultados do Henrique e meus..

%Observaçoes, reducao e medidas (ppxf e ifscube)

%resultados (descrever as figuras)

%Discussoes (kinemetry, taxa e potencia cinetica do outflow)

%fazer a razao entre taxa de outflow em H2 e HII como fiz no paper de 2023 com o NIFS

%PS: colocar curvas de contornos...

%Bora, paper!

Active Galactic Nuclei (AGN) play a crucial role in shaping the evolution of their host galaxies by regulating star formation and influencing key properties such as stellar mass and chemical enrichment \citep{Harrison2017, Harrison2024}. This interaction is primarily driven by AGN feeding, the process in which gas fuels the supermassive black hole \citep[SMBH;][]{Storchi2019}, and AGN feedback, which manifests through radiation, jets, and gas outflows \citep{Kormendy2013, morganti2017, Harrison2018}. The outflows can coexist in the same galaxy as ionized, neutral atomic, and molecular gas, collectively known as multiphase outflows \citep{Cicone2018}.

The outflows observed in the ionized gas ($ T_{\rm gas} \sim 10^{4}$\,K) are usually traced by strong optical emission lines, such as  [O\,{\sc iii}]$\lambda5007$, which may exhibit broadened or asymmetric line profiles. In nearby AGNs, ionized outflows are typically restricted to the inner few kiloparsecs of the galaxies \citep{Revalski2018, Alice2019, RARiffel2023, Munoz2024, Gatto24}. However, in more luminous AGNs, they can span the entire host galaxy or even extend beyond it, reaching mass outflow rates of $\sim 1\,\rm M_\odot \,\mathrm{yr}^{-1}$ \citep{Guilin2013, Guilin2013b, Carniani2015, Dall'Agnol2021,Tozzi2024}. In contrast, the neutral atomic outflow ($T_{\rm gas} \sim 10^{2-3} \rm \,K$) can be traced by the H\,{\sc i} 21 cm and  Na I D$\lambda\lambda$5890,5896 absorption lines. These outflows can be fast ($>10^3$ km~s$^{-1}$) and are essential for gas cooling \citep{Morganti2013, Morganti2016, Rupke21}. 

The majority of the total outflowing mass is likely carried in the molecular gas phase. Molecular outflows are typically denoted as hot, warm, or cold, depending on the excitation temperature of the transitions used to observe them. The hot molecular outflow component ($T_{\rm gas} \gtrsim 10^{3}$\,K) is observed in the NIR through ro-vibrational H$_2$ emission lines. This component is not common and when detected, represents only a small fraction of the outflow mass, with typical mass outflow rates of only 10$^{-5}$ to 10$^{-2}$  M$_\odot$\,yr$^{-1}$ measured for nearby AGN \citep[e.g.][]{RARiffel2015, RARiffel2021, RARiffel2023, Bianchin2022, Ceci2025, Ulivi2025}. The warm molecular outflow ($10^2 \lesssim T_{\rm gas} \lesssim 10^3 $\,K) is detected in the mid-infrared (MIR) via rotational H$_2$ emission lines and can exhibit a mass outflow rate higher than that of the ionized phase \citep{Holden2023, CostaSouza2024}. The cold molecular outflow ($T_{\rm gas} \sim 10^2$\,K) is observed in the far-infrared (FIR) and (sub)millimeter wavelengths through OH, CO, and HCN transitions, for instance. These outflows can be fast ($>600$ km~s$^{-1}$) and represent the dominant mass component of the outflow in powerful AGNs \citep{Morganti2015, Veilleux2020}, whereas in nearby AGNs, they may account for only a small fraction of the total outflow mass \citep{DallAgnol2023}.

Molecular hydrogen is typically associated with star formation processes in galaxies \citep{Omont2007}, however, some galaxies exhibit an excess of H$_2$ emission that cannot be fully explained by the radiation from young stars alone \citep{Ardila2004, Ardila2005,Zakamska2010, RRiffel2013, Motter2021}. These galaxies show enhanced H$_2$ line intensities in their spectra relative to what is expected from star-formation, suggesting the presence of additional excitation mechanisms, such as shock heating from supernovae or AGN outflows \citep{Ogle2010, Lambrides2019, RARiffel2020, Ogle2025}. Alternatively, some studies also attribute this excess to merger events \citep{Ellison2015,Larson2016,Petric2018}. 

%One galaxy that exhibits a multiphase outflow and an excess of H$_2$ emission is the nearby \citep[$z\simeq$ 0.0450;][]{Sandage1966} radio galaxy 3C\,293 (UGC\:8782).

The radio galaxy 3C 293 (UGC\:8782) for instance, is a nearby \citep[$z\simeq$ 0.0450;][]{Sandage1966} example of the fore mentioned group. This galaxy hosts an AGN classified as a Low Ionization Nuclear Emission Region \citep[LINER;][]{Véron2006} and is considered a restarted radio source, indicating that it has resumed nuclear activity after a period of quiescence, with evidence of multiple epochs of activity \citep{Saripalli2007, Joshi2011, Kukreti2022}. The host galaxy is classified both as spiral \citep{Sandage1966, deVancoulers1991} and dusty elliptical \citep{Ebneter1985, Tremblay1985}, with its rich interstellar medium (ISM) attributed to a merger event between an elliptical and a spiral galaxy. This merger likely led to the high concentration of molecular gas in the inner region, which may be fueling the AGN \citep{Evans1999, Floyd2006}. 

As a radio-loud source belonging to the Fanaroff-Riley II (FR II) class \citep{Fanaroff1974, Liu2002}, 3C\,293 exhibits a double-double radio structure, with inner radio lobes extending over $\sim$ 4 kpc \citep{Machalski2016}. Observations reveal a complex outflow system, including a neutral gas outflow driven by the  radio jet's interaction with the ISM \citep{Morganti2003}, and an ionized outflow on the opposite side, where shocks are generated also by jet-ISM interaction \citep{Emonts2005, RARiffel2023b}. 

Additionally, \cite{CostaSouza2024} detected a warm molecular outflow using James Webb Space Telescope (JWST) Medium Resolution Spectrometer (MRS) of the Mid-InfraRed Instrument (MIRI). This warm outflow has a mass-outflow rate of one order of magnitude larger than that of the ionized outflow, reaching a maximum outflow rate of $4.9 \pm 2.04 ~\rm M_{\odot} yr^{-1}$ at $\sim 900 \rm ~pc$ from the nucleus, and a kinetic power of $1.26 \times 10^{42} \rm \,erg s^{-1}$, being capable of expelling the warm molecular gas from the inner region of the galaxy in $\sim 1 \rm \,Myr$. \cite{Labiano2014} conducted interferometric observations of the $\rm ^{12}CO(1-0)$ and $\rm ^{12}CO(2-1)$ lines in search for a cold molecular gas outflow. However, no signatures of such an outflow were detected in this galaxy. Nevertheless, the data revealed the presence of a large disk of cold molecular gas, with regular circular rotation around the nucleus \citep{Labiano2014}.  

The excess of $\rm H_2$ emission in 3C\,293 is probably produced by shocks, as suggested by multiple studies. \cite{Ogle2010} argue that the radio jet can produce the observed $\rm H_2$ emission by heating the ISM and generating shocks that propagate into molecular clouds, which in turn excite the $\rm H_2$. This mechanism is further supported by \cite{Guillard2012}, who found that the $\rm H_2$ emission cannot be explained only by stellar UV radiation or AGN X-rays, but rather by the dissipation of kinetic energy from the radio jet through shocks. \cite{RARiffel2025}, using MIRI MRS observations of 3C\,293, also associate the excess of $\rm H_2$ with shock heating, likely caused by the interaction of the radio jet with the ISM. This interaction, co-spatial with the radio core, drives both warm molecular and ionized outflows.

%The investigation of multiphase outflows in galaxies like 3C\,293 is crucial for understanding the mechanisms driving this phenomena, as well as their impact on the evolution of the host galaxy and its AGN. While previous observations have revealed ionized, neutral, and warm molecular outflows in this source, the hot molecular gas phase has not yet been explored. The high resolution of the JWST  Near-Infrared Spectrograph (NIRSpec) now provides a new opportunity to complete our understanding of the multiphase outflows in 3C\,293.

In this study, we use JWST Near-Infrared Spectrograph (NIRSpec) to complete our understanding of the multiphase outflows in 3C\,293. We investigate the dynamics of the ionized gas, traced by the [Fe\,{\sc ii}]$ \rm \, 1.64\, \mu m$ and $\rm Pa\alpha$ emission lines, and hot molecular gas, traced by the $\rm H_2 \, 2.12 \, \mu m$ line, as well as the stellar component, in the inner region of the galaxy. This paper is organized as follows. Section \ref{sec:data} describes the observations, data reduction procedures, and analysis methods. In Section \ref{sec:results}, we present our main results, which are then discussed in Section \ref{sec:Discussion}. Finally, in Section \ref{sec:conclusions} we summarized our main conclusions.

\section{Data and Measurements} \label{sec:data}

\subsection{Observation and data reduction}
The galaxy 3C\,293 is part of the Blowing Star Formation Away in AGN Hosts (BAH) project \citep{CostaSouza2024, RARiffel2025}, which aims to investigate the molecular gas properties in the central regions of nearby AGN hosts exhibiting strong H$_2$ emission line. The observations were conducted using JWST’s NIRSpec \citep{Jakobsen2022, Gardner2023, Boker2023} in Integral Field Unit (IFU) mode (Program ID: 1928; PI: Riffel, R. A.), employing the G235H/F170LP grating-filter combination, which covers a wavelength range of $\rm \sim 1.66 - 3.17 \,\mu m$. This configuration provides a resolving power of $\rm R \sim 2700 $, which corresponds to an instrumental full width at half maximum (FWHM) of $8.89$ \AA \ at $\lambda = 2.4 \,\mu m$, or a instrumental broadening of $\sigma_{\rm inst} \simeq 47 \, \rm km \, s^{-1}$.  A 6-point small-cycle dither pattern was applied, with the NRSIRS2RAPID readout mode, and an integration of 40 groups per exposure. Additional exposures intended for leakage correction were acquired but covered only a single dither position and used only to identify outliers. Any subtraction was performed during reduction due to the absence of significant variations in final data. 

Data processing was carried out using version 1.12.5 of the JWST Science Calibration Pipeline \citep{bushouse_2024}, employing the \texttt{jwst\_1183.pmap} reference configuration and following the recommended reduction steps. During the \texttt{Detector1} stage, basic detector-level corrections and slope fitting were applied. To mitigate artifacts from hot pixels, we applied a custom flagging routine inspired by \citet{Pontoppidan22}, which detects outlier pixels based on their intensity and spatial context. The cleaned rate files were then processed through the \texttt{Spec2} stage, with pixel replacement enabled. No pixel-level or master background subtraction was performed due to the lack of dedicated background exposures. To suppress inverse frequency noise (1/f), we applied the NSClean algorithm \citep{nsclean_2024}. In the \texttt{Spec3} stage, we adjusted outlier detection parameters (\texttt{kernel="1 11"}, \texttt{threshold=99.5}) and generated combined data cubes from both detectors with 0.1 arcsec spaxels by setting \texttt{output\_type="multi"}. The \texttt{.crf} files produced in this step were used as input for a custom artifact rejection script, which flagged residual outliers by identifying spatial intensity contrast patterns. A final run of the \texttt{Spec3} stage was performed with these updated masks, resulting in the rejection of approximately 1-2\% of pixels.

Given the spatial undersampling inherent in NIRSpec IFU data, individual spaxel spectra often display low-amplitude oscillations known as “wiggles” \citep[e.g.][]{Perna23,Law23}. To alleviate this, we employed a resolution-smoothing approach in which each spaxel’s spectrum was replaced with a version derived from the surrounding spectra, normalized to preserve its continuum. This method, akin to applying a wavelength-slice Gaussian blur, helps suppress wiggles while maintaining flux conservation. While this technique can cause slight variations (typically 5-10\%) in the flux of emission lines, it significantly improves spectral quality. Such “wiggle mitigating” step is essential to obtain the stellar kinematics in the regions closer to the nucleus of the galaxy. 

Our final data cubes have a spatial resolution of $\sim 0.13 $ arcsec at $2.4 \,\mu m$, as given by the FWHM of the JWST Point Spread Function (PSF) \citep{Eugenio2024}. At the distance of 3C\,293, it corresponds to a physical scale of $\sim 130 $ pc, assuming $\rm H_0 = 67.8 \, km\, s^{-1} \,Mpc^{-1}$ and $\Omega_{\rm matter} = 0.308$. 

%Our final data cubes have a spatial resolution of $\sim 0.32 $ arcsec, as given by the FWHM of the JWST Point Spread Function (PSF), measured from datacube for the star HD 2811. At the distance of 3C\,293, it corresponds to a physical scale of $\sim 300 $ pc.

% EM CONSTRUÇÃO
%The observations were conducted with JWST NIRSpec G235H/F170LP, which covers a wavelength range of $\rm \sim 1.66 - 3.17 ~\mu m$, with a resolving power $\rm {R = 2700}$, resulting in an instrumental dispersion of ${\sigma_{\rm inst} \simeq 47 \rm ~km ~ s^{-1}}$. 

%Consequently, after applying the stellar templates to derive the kinematics, it was necessary to correct the velocity dispersion values obtained from the {\sc ppxf} due to the resolution mismatch. Additionally, the stellar velocities were adjusted by accounting for the systemic velocity of the galaxy, which was determined to be $\rm V_{\rm sys} = 13525 \rm ~km~s^{-1}$. 

\subsection{Stellar kinematics}\label{subsec: sk}

\begin{figure*}[t]
    \centering
    \includegraphics[width=\linewidth]{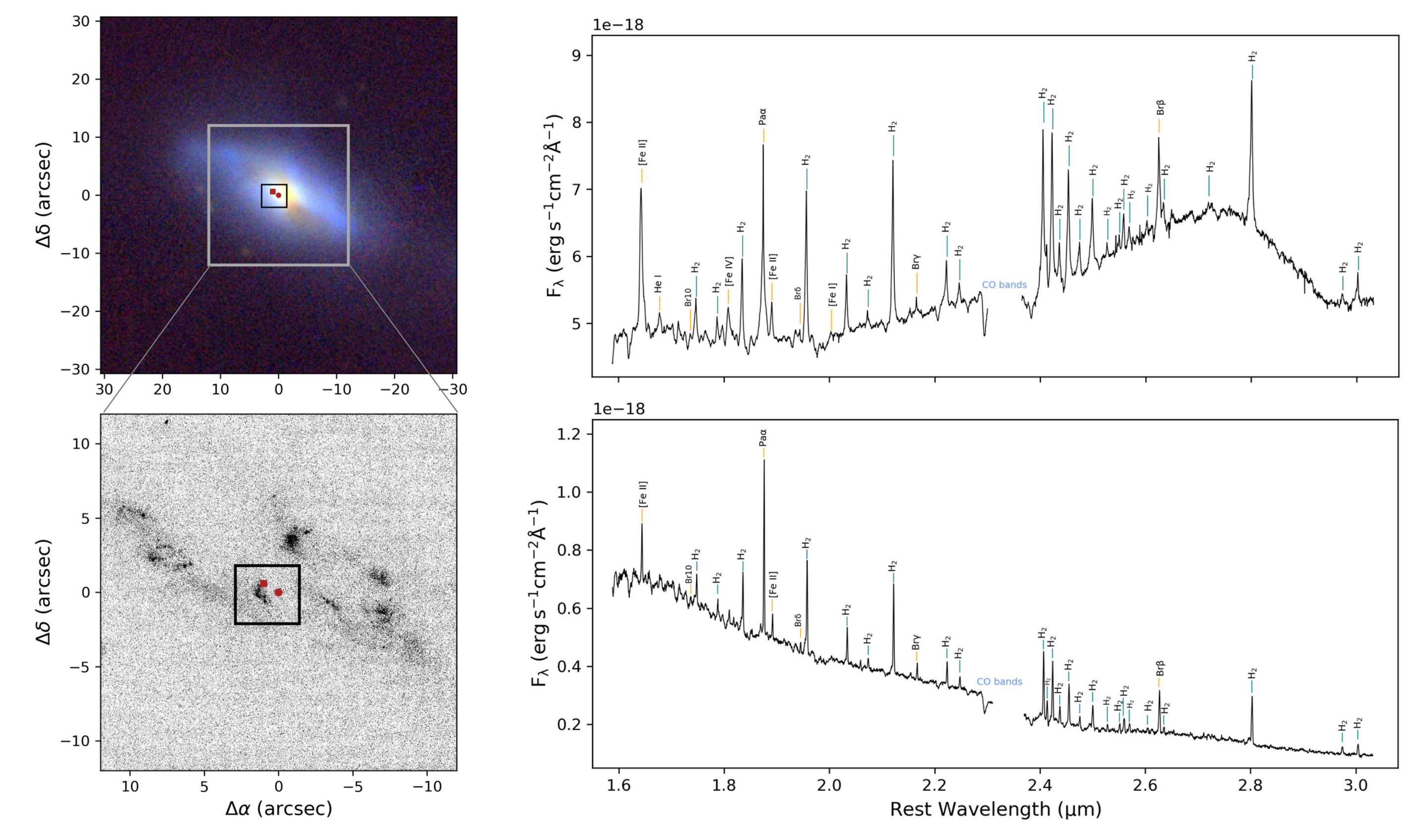}
    \caption{\textbf{\textit{Top left:} Composite image of 3C\,293 (UGC 8782) from Pan-STARSS data archive \citep{Chambers2019, Flewelling2020} in the \textit{y} (9633 \AA), \textit{i} (7545 \AA), and \textit{g} (4866 \AA) bands. The black rectangle marks the JWST/NIRSpec FoV, while the gray rectangle indicates the FoV of the Hubble Space Telescope (HST) near-ultraviolet (NUV) Cs2Te Multianode Microchannel Array (MAMA) detector. \textit{Bottom left}:  STIS NUV MAMA image of 3C\,293 \citep{Allen2002}, with the JWST/NIRSpec FoV indicated by a black rectangle. The red circle marks the IR nuclear position, and the red square marks an extranuclear position. \textit{Top right:} Nuclear spectrum of 3C\,293, obtained at the position of the continuum peak, corresponding to the red circle in the images. \textit{Bottom right:} Extranuclear spectrum of 3C\,293, observed with JWST/NIRSpec, at the location of the red square in the images. The orange lines highlight the ionized emission lines, while the teal lines indicate the molecular ones. The position of the CO absorption bandheads is also identified.}}
    \label{fig:spectra}
\end{figure*}

One approach to investigate the AGN's influences on the ISM is to determine whether the gas moves in the same way as the stars and whether this motion can be linked to the gravitational potential of the galaxy. This can be done by mapping the stellar kinematics in the inner region of the host galaxy. To achieve this, we used the penalized pixel-fitting ({\sc ppxf}) method \citep{Cappellari2004, Cappellari2017, Cappellari2023}, which determines the best fit to a galaxy spectrum by convolving it with stellar templates spectra. These templates are used to reproduce the galaxy's spectrum under the assumption that the line of sight velocity distribution (LOSVD) of the galaxy can be reproduced by a Gaussian function or by Gauss-Hermitte series.

In our observations, the JWST detector gap falls within the region of CO absorption bands at $\sim\,$2.3\:$\mu$m, which are typically prominent in nearby galaxies and provide important constraints for measuring stellar kinematics.  The position of this gap varies in wavelength across spaxels, affecting the analysis of this spectral feature. To mitigate this issue, we opted to used the {\sc ppxf} method across the full spectral range to derive the stellar kinematics, rather than relying solely on CO bands region for kinematic tracing.

We used the E-MILES stellar population templates \citep{Vazdekis2016}, which represents an improvement over the original MILES templates \citep{SanchezBlazquez2006} by extending coverage beyond the visible range. The E-MILES templates combine MILES with other empirical libraries, such as Indo-US \citep{Valdes2004}, CaT \citep{Cenarro2001a, Cenarro2001b}, and IRTF \citep{Cushing2005, Rayner2009}. E-MILES models are reliable for stars with ages $>\,$ 30 Myr and has a constant spectral resolution of $\sigma \simeq 60 \rm \, km\, s^{-1}$ in the NIR. The spectral fitting was performed using additive Legendre polynomials to match the continuum shape, along with an assumption of a Gaussian LOSVD, and with the \textit{clean} parameter enabled to reject outliers.  

The {\sc ppxf} method provides information on the stellar velocity ($ \rm v_{\star}$) and velocity dispersion ($\sigma_{\star}$) for each spaxel, along with their corresponding uncertainties.

\subsection{Fitting of the emission-line profiles}\label{subsec:gaussian_fit}

The composite image of 3C\,293 (UGC 8782), obtained from the Pan-STARRS data archive \citep{Chambers2019, Flewelling2020}, is shown in the top left of Figure~\ref{fig:spectra}, with the JWST/NIRSpec field of view (FoV) marked in black. The FoV of the Hubble Space Telescope (HST) near-ultraviolet (NUV) Cs2Te Multianode Microchannel Array (MAMA) detector is indicated by a gray rectangle. The STIS NUV MAMA image of 3C\,293 \citep{Allen2002} is shown in the bottom left, with the NIRSpec FoV marked by a black rectangle. In the right panel, we present the corresponding spectra extracted from the datacube \textbf{at the nuclear position (top right), which corresponds to the position of the continuum peak and is marked with a red circle, and at an extranuclear position (bottom right), indicated by a red square in the images. Both spectra were obtained using a single-spaxel aperture. The extranuclear spectrum was obtained at the position (1.0, 0.6) arcsec relative to the nucleus, chosen to represent a typical spectrum from the disk region of the galaxy}. 
 The spectra display numerous molecular emission lines (teal), ionized gas emission lines (orange), CO absorption bandheads, and other stellar features.

To obtain flux and kinematic information about the ionized and hot molecular gas in the inner region of 3C\,293, we used the Python package {\sc ifscube} \citep{RuschelDutra2021}. The emission-line profiles are fitted on the original datacube, without the subtraction of the stellar population contribution, as is commonly done in near-infrared studies \citep{RARiffel2023,Delaney25}, since in this spectral range the underlying stellar absorption and continuum features are weak and have a negligible effect on the derived line fluxes; additionally, the spectral region beyond 2.4~$\mu$m includes dust and H$_2$O ice features \citep{Perna24,Donnan2024} not accounted for in stellar templates, further justifying this approach. This package enables us to fit emission lines by Gaussian functions and model the continuum using a polynomial function for each spaxel in the data cube. The fitting process begins at the nucleus, identified by the spaxel with the peak continuum emission. The best-fit parameters from this spaxel are then used as initial guesses for the surrounding spaxels, following a looped path through the cube. A fifth-order polynomial provided the best representation of the continuum. Additionally, we enable the \textit{refit} parameter, which uses the best-fit parameters from spaxels located within 0.3 arcsec as initial guesses for subsequent fits. 

To trace the kinematics of the ionized gas, we selected two emission lines with high signal-to-noise ratios (S/N): [Fe\,{\sc ii}]$ \rm \,1.64\, \mu m$, a shock indicator originating from partially ionized regions \citep{StorchiBergmann2009, RRiffel2013, Zakamska2014b, Colina2015, Motter2021, Calabro2023}, and $\rm Pa\alpha$, which arises from fully ionized gas. For molecular gas kinematics, we used the $\rm H_2 \, 2.12\, \mu m$ emission line, as it is one of the strongest lines in the near-infrared.  The  $\rm H_2$ emission line was well-fitted with two Gaussian components, a narrow one representing disk rotation and a broad one tracing outflow features. The [Fe\,{\sc ii}] and $\rm Pa\alpha$ required three Gaussian components - a narrow disk component, a broad outflow component, and an additional broader component representing a faster outflow phase. This is a similar procedure as used in \citet{CostaSouza2024} in the analysis of the JWST/MIRI observations of 3C\:293. 

\begin{figure}
    \centering
    \includegraphics[width=0.9\linewidth]{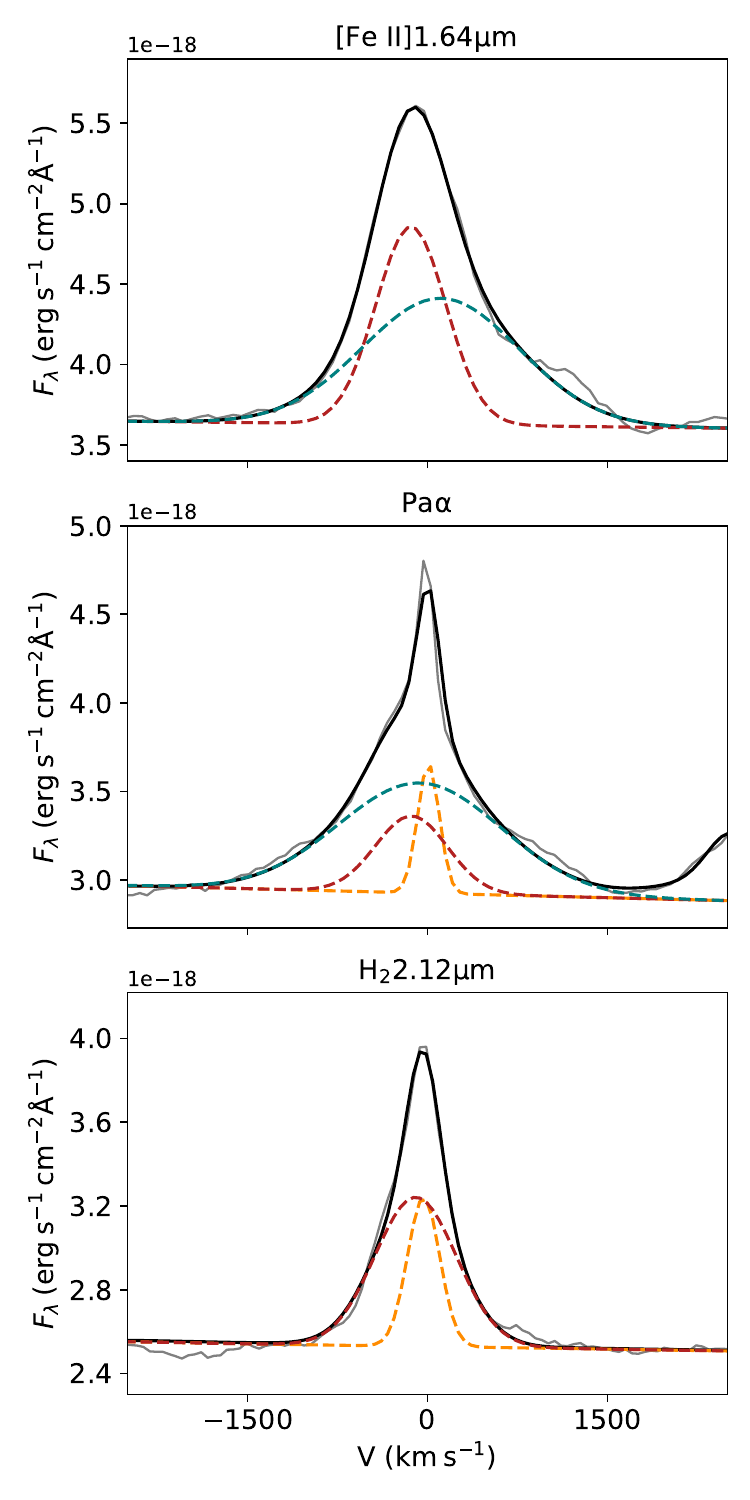}
    \caption{Examples of fits of the [Fe\,{\sc ii}]$\,1.64\:\mu$m (top), Pa$\alpha$ (middle) and H$_2\,2.12\,\mu$m (bottom) emission-line profiles for the nuclear spaxel. The observed profiles are shown as thin gray  lines and the best-fit models are shown in black. The narrow component is represented by dashed orange lines, the broad component by dashed dark red lines, and the very broad component by a dashed teal line. }
    \label{fig:components}
\end{figure}

Fig.~\ref{fig:components} presents the fit of [Fe\,{\sc ii}]$\,1.64\:\mu$m (top), Pa$\alpha$ (middle) and H$_2\,2.12\,\mu$m (bottom) emission-line profiles for the nuclear spaxel, performed using the {\sc ifscube} Python package. The observed profiles are shown as thin gray lines. The narrow, broad, and very broad Gaussian components are represented by orange, dark red, and teal dashed lines, respectively, while the resulting composite fit is shown in black. No narrow component was required to fit the [Fe\,{\sc ii}]$\,1.64\:\mu$m profile in the nuclear spaxel, indicating that the broad (potentially associated with outflows) components dominate at the nuclear region of the galaxy.

The {\sc ifscube} code generates a data cube containing the best-fit parameters, which we then used to map the flux distribution and kinematics of the emission lines.

\begin{figure}
    \centering
    \includegraphics[width=\linewidth]{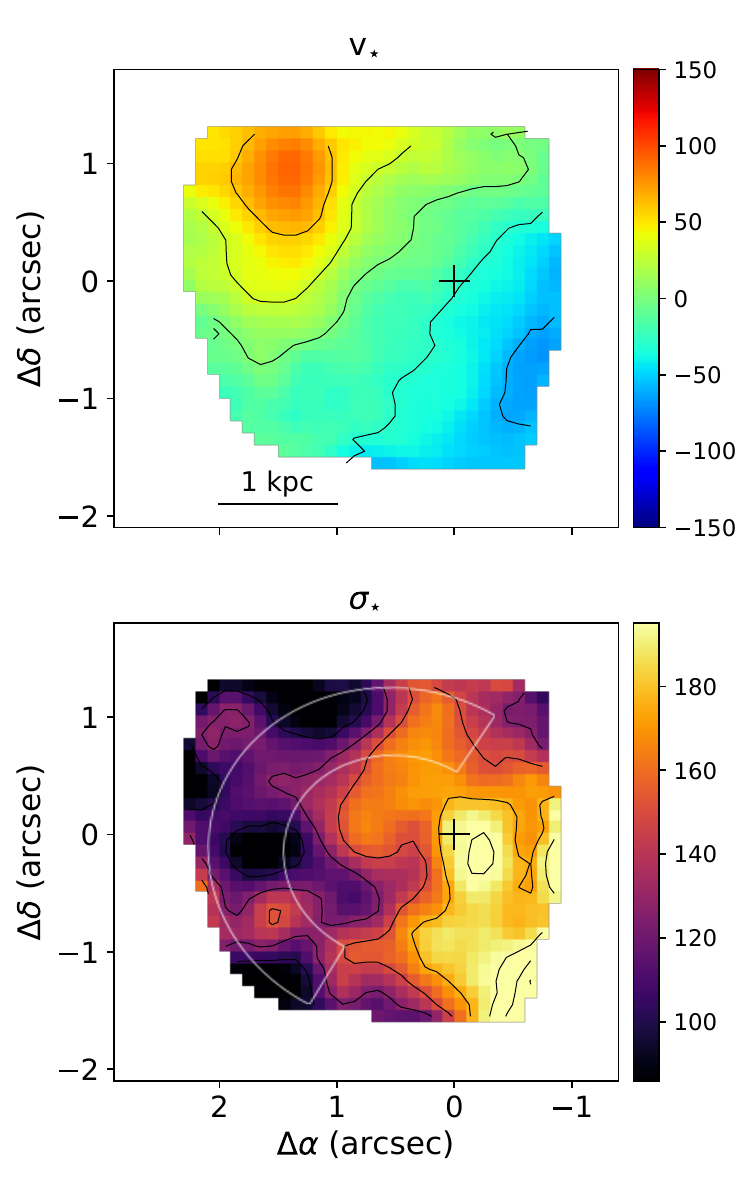}
    \caption{The top panel displays the stellar velocity field (in units of $\rm km~s^{-1}$) corrected for systemic velocity, while the bottom panel shows the stellar velocity dispersion (in units of $\rm km~s^{-1}$), corrected for instrumental broadening. The white partial ring indicates a region of lower velocity dispersion, attributed to an intermediate-age stellar population. All panels are oriented with north up and east to the left. The cross indicates the location of the NIR nucleus, which corresponds to the peak of the continuum emission. Regions with uncertainties larger than 20\:km\:s$^{-1}$ are masked.}
    \label{fig:stellar}
\end{figure}

\begin{figure*}[t]
    \includegraphics[width=\linewidth]{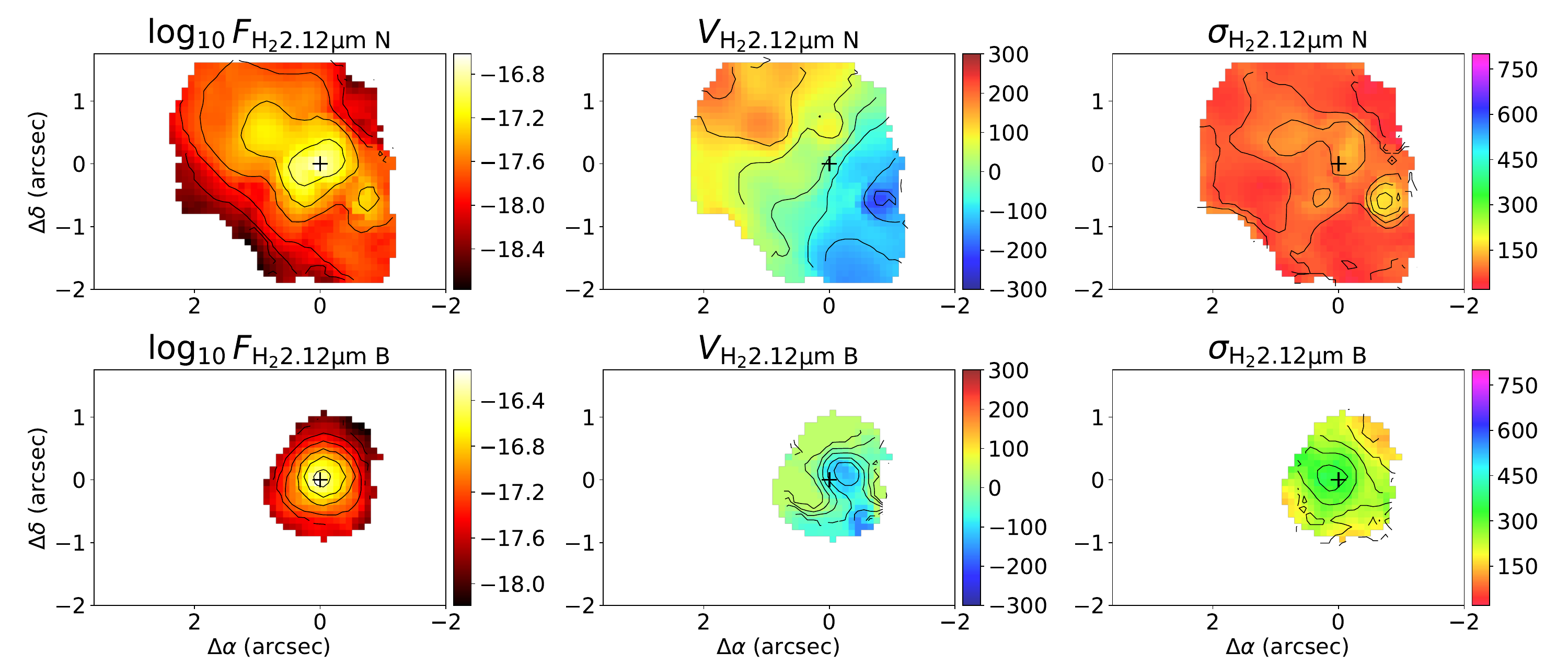}
    \caption{The top row shows the flux distribution (in units of $\rm erg\, s^{-1} cm^{-2}$, logarithmic scale), velocity (in units of $\rm km ~s^{-1}$), and velocity dispersion (in units of $\rm km ~s^{-1}$) for the narrow component of the $\rm H_2\, 2.12 \,\mu m$ emission line. The second row displays the same maps for the broad component. The measurements are restricted to areas with at least $\rm 3\sigma$ detections above the continuum noise. The cross marks the nucleus, corresponding to the position of the continuum emission peak.}
    \label{fig:H2}
\end{figure*}

\begin{figure*}[t]
    \includegraphics[width=\linewidth]{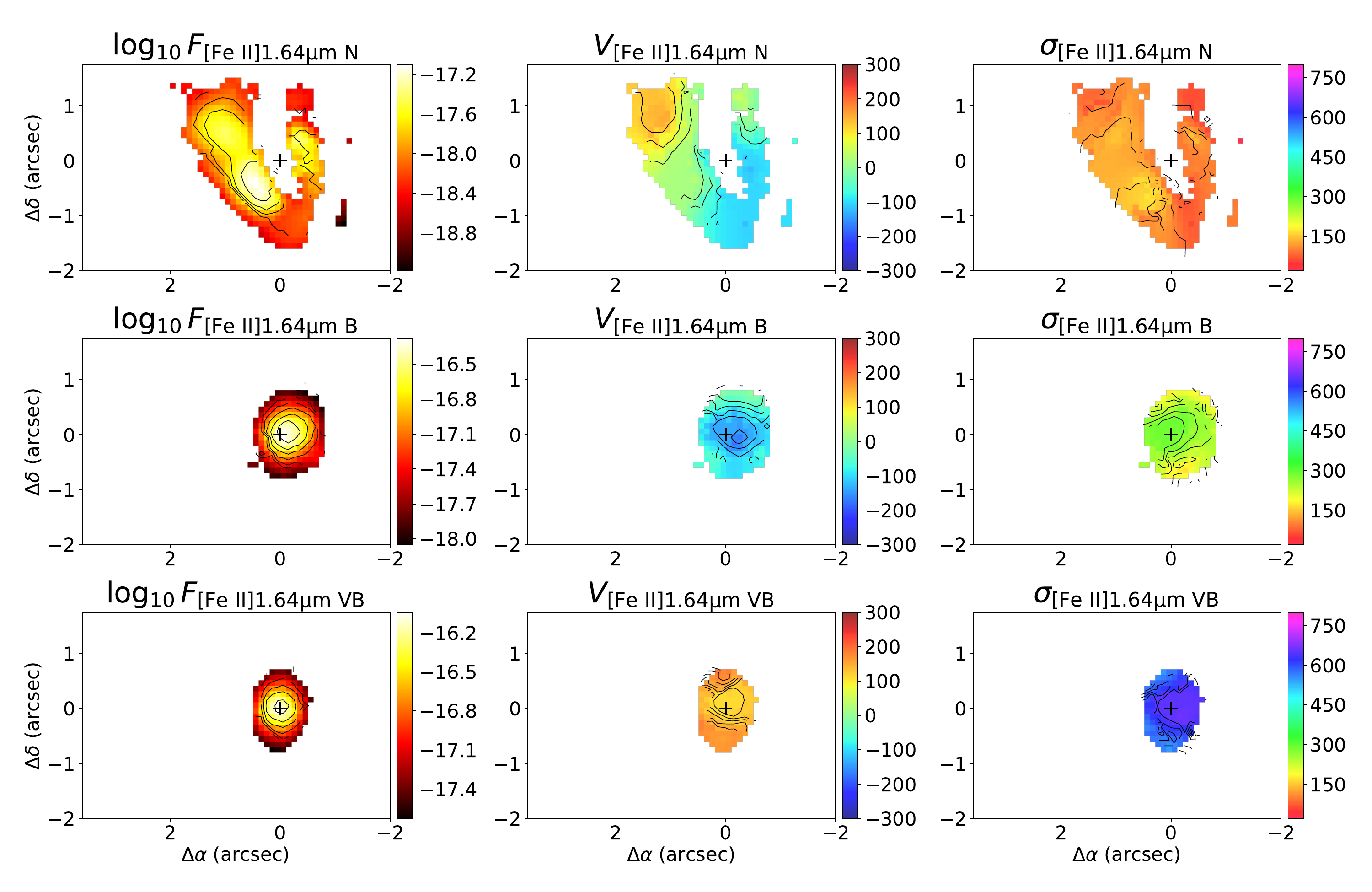}
    \caption{Same as Fig.~\ref{fig:H2}, but for the  [Fe\,{\sc ii}]$\rm 1.64 \mu m$ emission line. In addition, the third row shows the corresponding maps for the very broad component.}
    \label{fig:FeII16}
\end{figure*}

\begin{figure*}[t]
    \includegraphics[width=\linewidth]{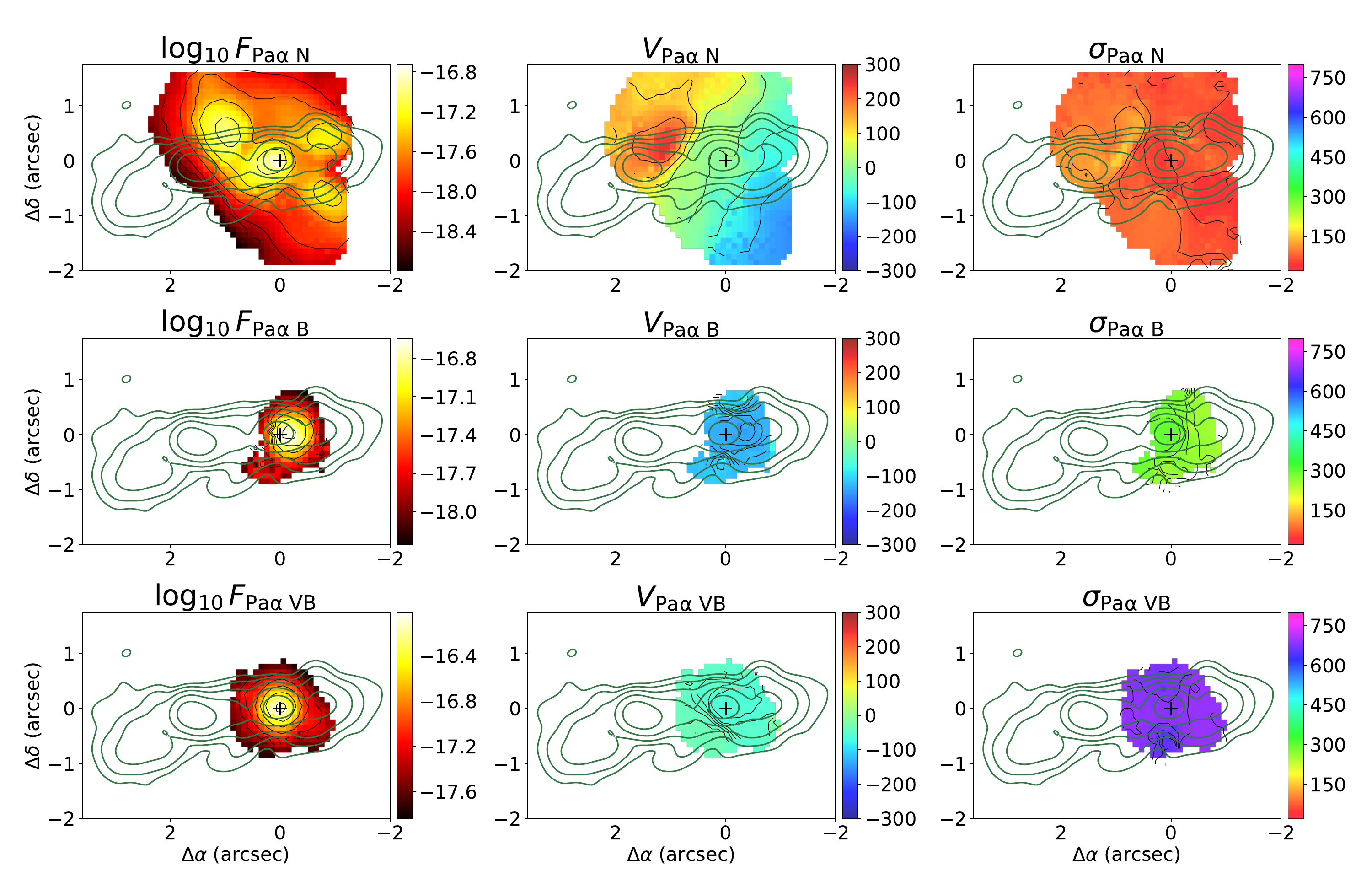}
    \caption{Same as Fig.~\ref{fig:H2}, but for $\rm Pa\alpha$. In addition, the third row shows the corresponding maps for the very broad component.  The green contours are from the 120–168 MHz radio continuum emission from \cite{Kukreti2022}}
    \label{fig:Pa}
\end{figure*}

\section{Results} \label{sec:results}

\subsection{Stellar kinematics}

Fig.~\ref{fig:stellar} presents the stellar velocity ($ \rm v_{\star}$)  and velocity dispersion ($\sigma_{\star}$) maps, obtained through the {\sc ppxf} method. The spatial scale is shown in the maps, the central cross represents the location of the peak of the continuum emission (hereafter NIR nucleus) and the white regions correspond to masked locations where the uncertainties in  $ \rm v_{\star}$ or $\sigma_{\star}$ are larger than 20 km\:s$^{-1}$. 

The velocity field has been corrected for systemic velocity ($V_{\rm sys} = 13\:412 \rm ~km~s^{-1}$), determined by fitting the observed velocity field by a rotation disk model, as will be discussed later in Sec.~\ref{subsec:diskprop}. The resulting map reveals a well-defined rotation pattern consistent with a disk-like motion, with blueshifted velocities in the southwest and redshifted velocities in the northeast, with a projected velocity amplitude of $\sim 100 \,\rm km\:s^{-1}$. 

The velocity dispersion map presents values in the range from $\sim 80 \,\rm km\:s^{-1}$ to $\sim 200\,\rm km \, s^{-1}$. The highest values are observed close to the peak of the continuum emission, represented as a cross in the map, and to southeast of it. Notably, we identify a partial ring structure approximately 1.5 arcsec east of the NIR nucleus, characterized by the lowest velocity dispersions (indicated by a white arch in Fig.~\ref{fig:stellar}). This feature likely traces intermediate-age stellar populations (0.3 - 0.7 Gyr), similar to those observed in other nearby active galaxies \citep[e.g][]{RARiffel2010, RRiffel2011, RARiffel2017, Diniz17}.  These populations were likely formed as a result of a merger event that occurred approximately 1 Gyr ago, triggering a burst of star formation \citep{Evans1999}. Indeed, this galaxy shows important fractions of intermediate-age stellar populations (Riffel et al., in preparation).

\subsection{Gas emission and kinematics}

In Fig.~\ref{fig:H2}, we present the flux and kinematic maps for the  $\rm H_2 \,2.12\, \mu m$ emission line.  The first row displays the flux, velocity, and velocity dispersion ($\sigma$) for the narrow component, attributed to the disk emission. The second row presents the measurements for the broad component, associated with the outflow. Only measurements where the amplitude of the corresponding emission-line component exceeds three times the standard deviation of the nearby continuum are included. The central crosses show the position of the NIR nucleus. The $\sigma$ maps are corrected for the instrumental broadening and the velocity fields are shown relative to the systemic velocity of the galaxy.

The flux distribution of the narrow component of $\rm H_2\,2.12\,\mu$m emission line extends across the entire NIRSpec FoV, with the emission peaking at the NIR nucleus and exhibiting an elongated structure along the northeast-southwest direction, aligned with the galaxy's major axis \citep[PA = 50$^\circ$;][]{Skrutskie2006}. Its velocity field exhibits a rotating disk pattern, similar to that observed for the stars, though somewhat more disturbed and with a higher projected velocity amplitude of $\sim200\,\rm km\,s^{-1}$. The $\sigma$ map presents overall low values, typically smaller than 100\,km\,s$^{-1}$, with a mean value of $\rm \sigma \simeq 80\,km\,s^{-1}$. The broad outflow component is detected within an inner radius of 1~arcsec, exhibiting a roughly round flux distribution and contributing about half of the  H$_2\,2.12\,\mu$m total flux. It is blueshifted relative to the narrow component, with velocities reaching up to $\sim300 \, \mathrm{km\,s^{-1}}$. The mean velocity dispersion is $\sigma \simeq 240 \, \mathrm{km\,s^{-1}}$, with the highest value of $\sigma \simeq 360 \, \mathrm{km\,s^{-1}}$ observed at the nucleus. Thus, the broad component represents the hot molecular phase of the outflows previously reported in the warm molecular \citep{CostaSouza2024} and ionized gas \citep{Emonts2005,RARiffel2023b}.

In Fig.~\ref{fig:FeII16}, we present the flux distribution and kinematics of each component of the [Fe\,{\sc ii}] $\rm 1.64 \, \mu m$ emission line, using the same approach applied in Fig.~\ref{fig:H2}. Similarly to the [Ar\:{\sc ii}]\:6.99\,$\mu$m emission in the mid-infrared observed with MIRI \citep{CostaSouza2024}, the [Fe\,{\sc ii}] and $\rm Pa\alpha$ display three kinematic components. In addition to the disk and outflow components (shown in the first and second rows, respectively) seen in the other emission lines, a ``very broad" component is also observed (shown in the third row), tracing a more turbulent phase of the ionized outflow. The flux map of the narrow [Fe\,{\sc ii}] component shows a distribution similar to that of H$_2$, elongated along the galaxy's major axis. However, the narrow [Fe\,\textsc{ii}] component is not detected at the nucleus or to its north (resulting in a V-shaped morphology), in a region co-spatial with a prominent dust lane seen in optical images , likely due to dust extinction \citep[][see also Fig.~\ref{fig:extinction}]{Floyd2006}. The rotation pattern is also evident in the [Fe\,{\sc ii}] velocity field, although it appears somewhat more disturbed, particularly in the region close to the nucleus. Finally, the corresponding $\sigma$ map shows values higher than those observed for the narrow H$_2$ component, but remains relatively low overall, with a mean value of $\sigma \simeq 100\, \mathrm{km\,s^{-1}}$. As for the H$_2$, the broad [Fe\,{\sc ii}] component also exhibits a roughly round flux distribution, but with higher blueshifted velocities and larger velocity dispersion values, having a mean $\sigma \simeq 240\, \mathrm{km\,s^{-1}}$. Additionally, the [Fe\,{\sc ii}] very broad component -- interpreted as originating from a faster outflow -- exhibits a flux distribution that is more concentrated toward the nucleus. It is characterized by predominantly redshifted velocities and significantly higher velocity dispersion values, with a mean of $\sigma \simeq 620\, \mathrm{km\,s^{-1}}$. These characteristics suggests that the [Fe\,{\sc ii}] emission traces a more disturbed phase of the outflow compared to the hot molecular phase, consistent with its origin in partially ionized zones.

In Fig.~\ref{fig:Pa}, we also present the flux distribution and kinematics for each component of the $\rm Pa\alpha$ emission line, a tracer of fully ionized gas zones. The 120–168 MHz radio continuum emission from \citet{Kukreti2022} is overlaid in green in all panels. The flux distribution for the disk component presents emission elongated along the galaxy's major axis, but with emission knots, likely associated with star-forming regions. In Fig.~\ref{fig:spectra}, we present the NUV MAMA image of 3C\,293, which reveals numerous emission knots associated to star forming regions. Although the NIRSpec FoV is much smaller than that of the NUV image, the Pa$\alpha$ emission knots do not appear to be co-spatial with the NUV knots, which are observed mainly to the east of the nucleus within the NIRSpec FoV.
A possible explanation for this apparent discrepancy could be extinction, which may obscure the same star-forming knots traced by Pa$\alpha$ at NUV wavelengths. \citet{Floyd2006} reported that the region of bright NUV emission near the nucleus is in fact co-spatial with the eastern radio jet. They attributed this excess either by synchrotron emission or to inverse Compton emission from low-energy electrons, although NUV excess could also result from star formation activity triggered by the jet. Alternatively, the emission knots seen in Pa$\alpha$ could also be associated with emission from gas compressed by the radio jet.  A more detailed investigation of the stellar population properties in 3C\,293 will be presented in future work through stellar population synthesis.

The velocity field of the narrow component presents a well-ordered rotation pattern consistent with gas motion in the plane of the disk, exhibiting overall low velocity dispersion values with a mean of $\sigma \simeq 80\, \mathrm{km\,s^{-1}}$. The broad outflowing component is co-spatial with the radio core, which is coincident with the NIR nucleus, and exhibits similar properties to those observed in other emission lines, with a mean velocity of $\sim - 120 \, \rm km\,s^{-1}$ and a mean velocity dispersion of $\rm \sigma \simeq 270 ~km~s^{-1}$. The very broad component, also co-spatial with the radio core, shows a flux concentration near the nucleus, but exhibits lower velocities and significantly higher velocity dispersion compared to the other components. Its velocity is close to zero, while the velocity dispersion has a mean value of $\sigma \simeq 700\, \mathrm{km\,s^{-1}}$, similarly to values seen in [Ar\:{\sc ii}]\:6.99$\mu$m  using JWST MIRI data and denominated as ``fast ionized outflows" \citep{CostaSouza2024}.

The flux distributions of the broad and very broad components of the three emission lines are spatially resolved, with FWHMs ranging from 0.6 arcsec for the broad component of [Fe\:{\sc ii}] to 0.8 arcsec for the broad component of H$_2$, 4.6–6.2 times larger than the instrumental PSF. Combined with the correspondence between the kinematic components detected here and those observed in MIR lines by \citet{CostaSouza2024}, this confirms that the ionized and hot molecular outflows are spatially resolved.

In summary, our NIRSpec observations reveal three kinematic components of the ionized gas in 3C\:293: one dominated by gas motions within the galaxy's disk and two associated with outflows, previously identified at other wavelengths and in different gas phases. In the next section, we determine the properties of the disk and outflows, as well as discuss dust attenuation using the flux distributions observed for each component.

\section{Discussion}\label{sec:Discussion}

3C\:293 is a complex system that has undergone a  merger event, during which large amounts of gas were accreted and are now likely the AGN fuel \citep[e.g.][]{Floyd2006}. It is one of the galaxies with the strongest H$_2$ emission in the local Universe \citep[e.g.][]{Ogle2010} and features a restarted radio jet \citep[e.g.][]{Akujor96,Beswick02} alongside a complex dust structure, including prominent dust lanes in the central region \citep[e.g.][]{Floyd2006}. The optical and infrared nuclei are offset, likely due to the presence of dust that obscures the nucleus in optical observations \citep[e.g.][]{CostaSouza2024}. Outflows of ionized \citep{Emonts2005,Mahony2016,RARiffel2023b}, warm molecular \citep{CostaSouza2024}, and neutral \citep{Morganti2003,mahony13} gas have been reported, although no outflows have been detected in cold molecular gas \citep{Evans1999,Labiano2014}. Despite this, the galaxy harbors significant amounts of molecular gas in its central region    \citep[2.2$\times$10$^{10}$\,M$_\odot$;][]{Labiano2014}, rotating within the plane of the disk. These characteristics make 3C\,293 an excellent laboratory for studying the interplay between the AGN and the interstellar medium.

\subsection{Gas extinction}

\begin{figure*}
    \centering
    \includegraphics[width=\linewidth]{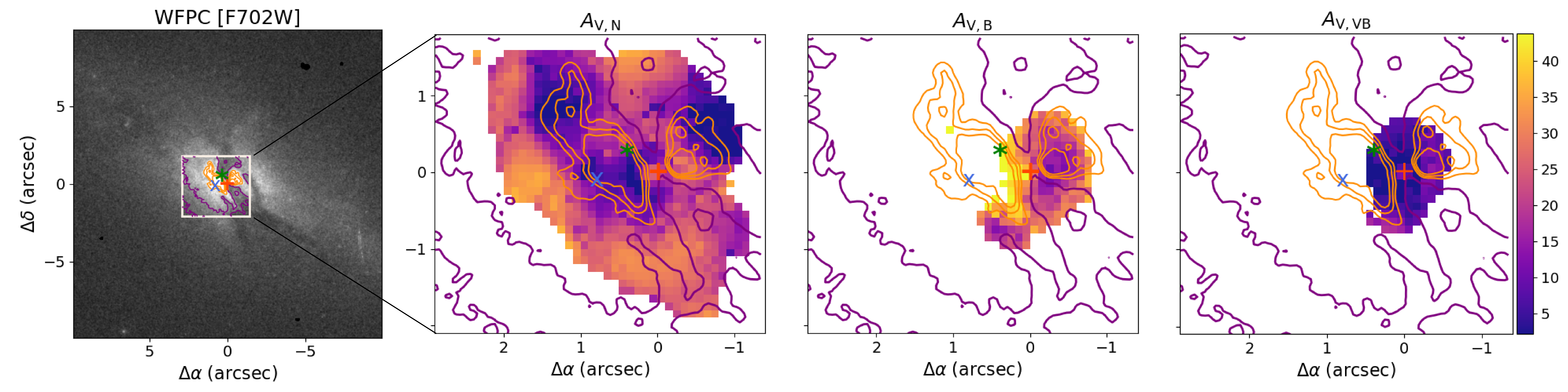}
    \caption{The first panel displays an HST (F702W) image taken with the Wide Field and Planetary Camera 2 (WFPC2), where the white rectangle marks the NIRSpec FoV. The second, third, and fourth panels show extinction maps derived from the line ratio between $\rm Br\beta/Pa\alpha$, corresponding to the narrow, broad, and very broad components, respectively. The red cross in each panel indicates the position of the nucleus, given by the peak of the continuum in the NIR, while the blue 'X' denotes the peak of the optical continuum. The green asterisk (*) represents the center of mass, as determined by the rotating disk model. The purple and orange contours overlaid on the HST image represent regions of higher (fainter regions) and lower (brighter regions) dust concentration, respectively. We superimpose these contours on the $A_V$ map to demonstrate the spatial correlation between the prominent dust lanes in the HST image and the peaks in our $A_V$ map.}
    \label{fig:extinction}
\end{figure*}

The galaxy 3C\,293 exhibits a complex dust morphology, characterized by five interlaced dust lanes extending over a scale of approximately 10 kpc. This structure likely results from the merger event that occurred around 1\,Gyr ago, as the dust orbits have not yet fully settled \citep{Koff2000, Ogle2010}. To study the spatial distribution of extinction in 3C\,293, we derived the visual extinction ($A_V$) using the $\rm Br\beta/Pa\alpha$ emission line ratio for each of the three kinematic components. The $A_V$ is given by

\begin{equation}
    A_{V} = \frac{-2.5}{q_{\rm Br\beta} - q_{\rm Pa\alpha}} \log \left[ \frac{(F_{\rm Br\beta}/F_{\rm Pa\alpha})_{\rm obs} }{(F_{\rm Br\beta}/F_{\rm Pa\alpha})_{\rm e}}  \right],
\end{equation} where $(F_{\rm Br\beta}/F_{\rm Pa\alpha})_{\rm obs}$ is the observed flux ratio, and $(F_{\rm Br\beta}/F_{\rm Pa\alpha})_{\rm e}$ = 0.13 is the emitted flux ratio, given by the theoretical intensity line ratio, assuming Case B H\,{\sc i} recombination in the low-density limit for an electron temperature of $\rm T_e = 10^4 \,K$ \citep{Osterbrock}. The parameters $q_{\rm Br\beta}= 0.074 $ and $q_{\rm Pa\alpha} = 0.133$ are obtained from the G23 extinction law \citep{G23a}, adopting $\rm R_v = 3.1$.  Replacing these values in the equation above, we obtain 

\begin{equation}
     A_{V} \simeq 42 \log \left[ \frac{(F_{\rm Br\beta}/F_{\rm Pa\alpha})_{\rm obs} }{ 0.13 }  \right].
\end{equation}

Fig.~\ref{fig:extinction} displays an archival Hubble Space Telescope (HST) image obtained with the Wide Field and Planetary Camera 2 (WFPC2) using the F702W filter. These observations were previously discussed in \citet{Floyd2006}, and the F702W image highlights the prominent dust lanes observed in 3C\,293. The white rectangle delineates the NIRSpec FoV. The remaining panels show the extinction maps derived from the observed line ratio between $F_{\rm Br \beta}$ and $F_{\rm Pa \alpha}$ in the V band for the narrow (disk), broad (outflow), and very broad (fast outflow) components, from right to left. To better illustrate the correspondence between dust features, contours of low (purple; dustier regions) and high (orange; brighter regions) emission from the HST image are overlaid on the $A_V$ map. The red cross in each panel marks the NIR nucleus, identified as the peak of the NIR continuum, while the blue `X' denotes the position of the peak of the optical continuum. The green asterisk (*) represents the position of the kinematical center, as determined by fitting the stellar velocity field by a rotating disk model, which is presented in Sec.~\ref{subsec:diskprop}. 

Notably, the maps for all components exhibit very high extinction values, with mean values of $A_V \simeq 19$ for the disk component, $A_V \simeq 27$ for the outflow component, and $A_V \simeq 7$ for the fast outflow component. The highest extinction value is observed for the outflow component, reaching up to $A_V \approx 35$, while the highest extinction values for the disk component are co-spatial with the dust lanes near the nucleus, consistent with the structure visible in the HST image (first panel). This spatial correlation is further emphasized by the superposition of the purple contours, from the HST image, onto the Av maps. 

Such very high extinction values are also observed for other AGN hosts. For instance, \cite{Ogle2025} also reported high $A_V$ values in Cygnus A, based on JWST NIRSpec observations, confirming previous measurements by \citet{Rogemar_CygA} using the Gemini Near-infrared Integral Field Spectrograph. In that case, a correlation was found between regions of highest $A_V$ and $\rm H_2$ knots, with extinction reaching 25 mag in these regions. In fact, dust plays a crucial role in the formation of $\rm H_2$, acting as a catalyst by providing surfaces for hydrogen atoms to interact and combine \citep{Wakelam2017, Grieco2023}. Other notable example of high extinction is Arp\:220 with $A_V$ values of up to 14 mag, as derived from NIRSpec observations \citep{Ulivi2025}. This highlights the power of infrared observations in probing deeper into the dusty circumnuclear regions of AGN hosts, revealing dusty multiphase outflows.

\begin{figure*}
    \includegraphics[width=\linewidth]{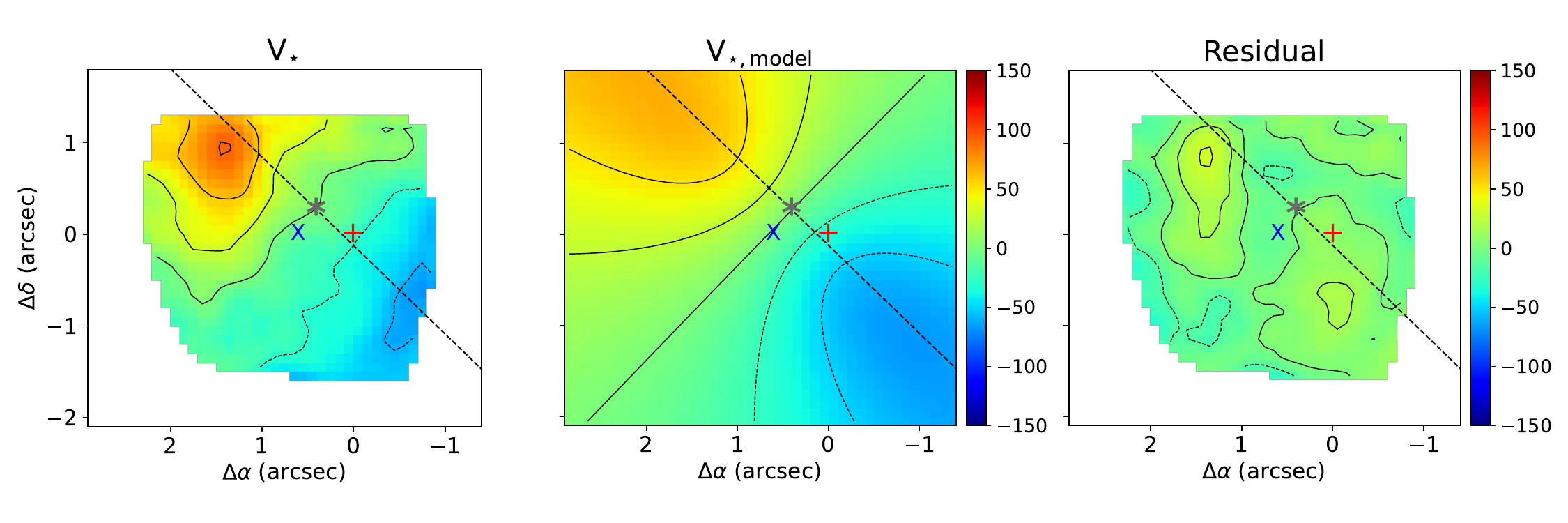} 
    \caption{The left panel displays the observed stellar velocity field, while the middle panel presents the best-fit rotating disk model and the right panel shows the corresponding residual map (observed - model). The dashed line indicates the orientation of the line of nodes. The kinematic center is indicated by a gray asterisk (*), while the red cross indicates the NIR nucleus, and the blue `X' shows the position of the optical nucleus. }
    \label{fig:rot_stars}
\end{figure*}

\subsection{Disk properties}\label{subsec:diskprop}

To determine the disk properties and identify non-rotational motions in the galaxy (e.g., outflows or inflows), we fitted the stellar and molecular gas velocity fields by a rotating disk model based on \cite{Bertola91}. This kinematic modeling was implemented using the Python package {\sc ifscube}. The model is described by

\begin{align}
    & V = V_{\rm sys} 
    \\ & + \frac{A R \cos(\Psi)\sin(\theta)\cos^p(\theta)}{\{R^2[\sin^2(\Psi) + \cos^2(\theta)\cos^2(\Psi)] + C_0 \cos^2(\theta)\}^{p/2}},  \notag
\end{align} where $V_{\rm sys}$ represents the systemic velocity of the galaxy, $A$ corresponds to the amplitude of the rotational velocity, $R$ denotes the radial distance of each pixel from the rotation center, and $\Psi = \psi - \psi_0$, with $\psi_0$ being the positional angle of each spaxel, defining the orientation of the line of nodes. The inclination of the disk is given by $\theta$. The parameter $p$ controls the slope of the rotation curve, ranging from an asymptotically flat rotation curve ($p=1$) to a system that has a finite mass ($p=3/2$). Finally, $C_0$ is the concentration parameter that adjusts the spatial scale of the velocity field.

We fitted the stellar velocity field using this model to determine both the disk parameters and the galaxy's systemic velocity, since stars are influenced solely by the gravitational potential and are unaffected by external perturbations (e.g., outflows). We also fitted the hot molecular gas disk, traced by the narrow component of the $\rm H_2 \, 2.12 \, \mu m$ emission line, in order to determine disk parameters and identify possible non-circular motions.  

\begin{figure*}
     \centering
     \includegraphics[width=\linewidth]{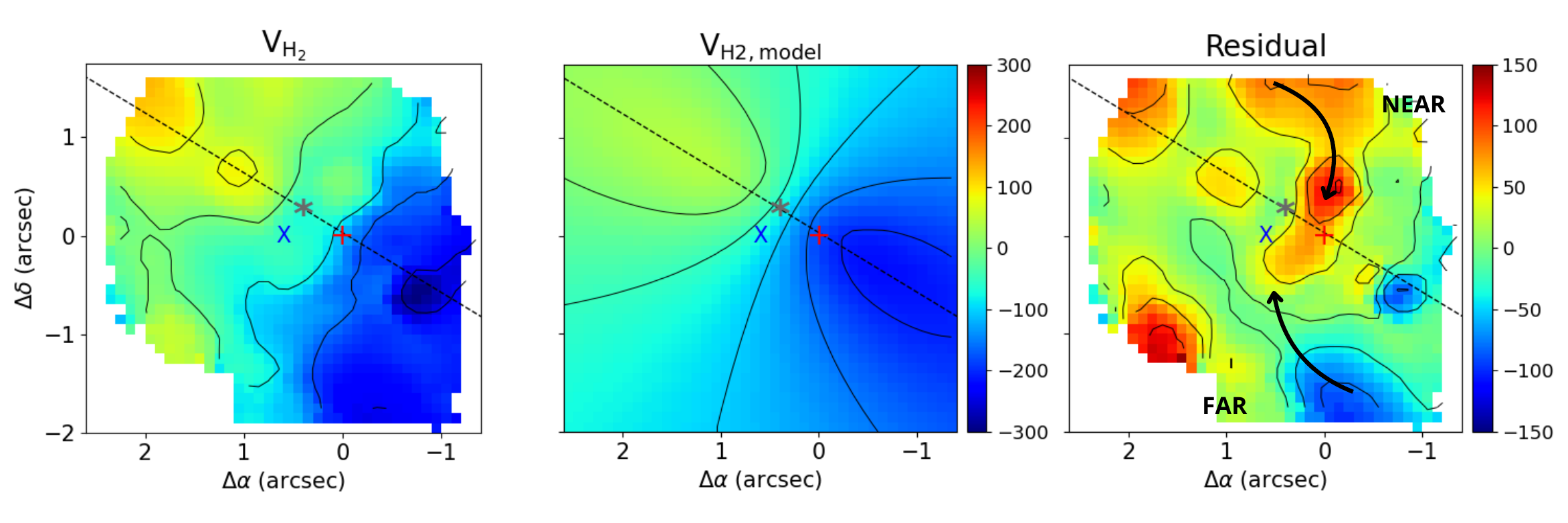} 
     \caption{Same as Fig.~\ref{fig:rot_stars}, but for the H$_2$\:2.1218\:$\mu$m emission line. The arrows highlight two high-velocity residuals tracing inflowing gas.}
     \label{fig:rot_h2}
\end{figure*}

In our stellar disk model, we treated the rotation center coordinates ($x_0$, $y_0$) as free parameters to determine the optimal dynamical center. The derived position is offset by approximately 0.2 arcsec north and 0.4 arcsec east relative to the NIR nucleus. We similarly allowed the remaining kinematic parameters to vary freely, obtaining the following values: $V_{\rm sys} = 13\:412\pm15 \rm ~km~s^{-1}$, $\psi_0 = 46^{\circ}\pm7^{\circ}$, $\theta = 53^{\circ}\pm6^{\circ}$, and $C_0 = 1.4\pm0.1$ arcsec.  The $p$ parameter was allowed to vary between 1.0 and 1.5, with the best fit converging to $p=1.5$. Fig.~\ref{fig:rot_stars} presents the observed velocity field (left), the best-fit model ($V_{\rm \star, model}$; middle), and the residual map ($V_{\rm \star} - V_{\rm \star, model}$; right). The kinematic center derived from the stellar model is marked by a gray asterisk (*), the NIR nucleus by a red cross, and the optical nucleus by a blue ‘X’.  The residual map present small values ($\lesssim20\:{\rm km\:s^{-1}}$) over all locations, indicating that the stellar kinematics are well described by pure rotation. The dashed line in each panel indicates the orientation of the line of nodes. The $\psi_0$ value is consistent with the kinematic and photometric position angles of the galaxy's major axis derived from optical integral field unit observations of the inner $3.4 \times 4.9\, \mathrm{kpc}^2$ of 3C\:293 \citep{RARiffel2023b}, as well as with the orientation of its large-scale disk, of $\sim50^\circ$ \citep{Skrutskie2006}.

Previous studies have reported spatial offsets between the optical nucleus and the radio core in 3C~293 \citep[e.g.,][]{Emonts2005, Mahony2016}. Furthermore, \citet{CostaSouza2024} identified a shift of approximately $0.6$ arcsec between the position of the peak of the mid-infrared continuum, observed with with MIRI (coincident with the radio core) and the optical nucleus from \citet{RARiffel2023}. This difference was attributed to higher dust attenuation in the optical data, as supported by the high extinction we find with NIRSpec (Fig.~\ref{fig:extinction}). In addition, we find that the galaxy's kinematic center is offset from both the optical and infrared nuclei. These discrepancies are likely a consequence of the merger event experienced by the galaxy \citep{Floyd2006}.

Following \citet{Diniz15}, we also modeled the molecular gas disk by fixing the parameters $V_{\rm sys}$, $\theta$, $p$, and the position of the kinematic center to the values obtained from the stellar velocity model. The remaining parameters were allowed to vary freely. This approach enables the determination of the maximum rotation velocity, which is generally higher for the gas than for the stars, as the gas is typically confined to a thin disk, whereas the stars exhibit greater velocity dispersion.  In this case, we derived the following best-fit values: $\psi_0 = 58^{\circ}\pm6^\circ$ and $C_0 = 0.8\pm0.1$ arcsec. Fig.~\ref{fig:rot_h2} shows the rotating disk model for the molecular gas, along with the observed velocity field and the residual map. The H$_2$ velocity field and corresponding model are asymmetric around the center, unlike the stellar velocity field. Such distortions could result from a recent merger \citep{Floyd2006}; however, the limited NIRSpec FoV prevents a more detailed investigation, for instance by comparing these deviations with dust structures on larger scales seen in HST images. The residual map shows higher values than those observed for the stars, reaching up to $130\,\mathrm{km\,s^{-1}}$. The largest residuals are mainly observed to the north in redshifts and to the south of the nucleus in blueshifts, approximately co-spatial with the dust lanes seen in the HST image (Fig.~\ref{fig:extinction}).

\subsection{Molecular gas inflows}

\begin{figure}
    \centering
    \includegraphics[width=0.8\linewidth]{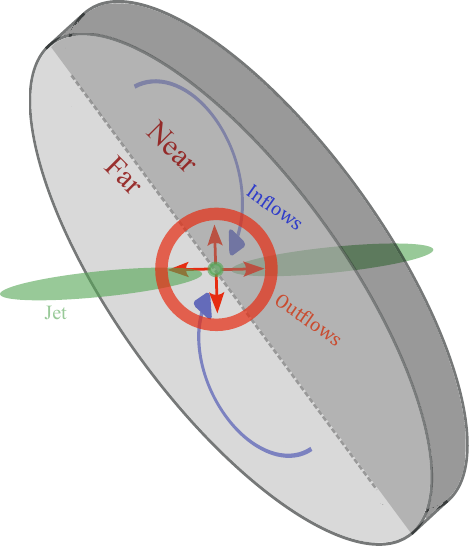}
    \caption{Schematic representation of the galaxy 3C\,293. The orientation of the galactic disk is shown, with the near and far sides labeled. Proposed inflows and outflows are indicated by blue and red arrows, respectively. The inner jet is illustrated in green.}
    \label{fig:schematic}
\end{figure}

 The eastern side of the radio jet in 3C\,293 is approaching us \citep[e.g.][]{Beswick04}, and a higher dust obscuration is observed in the northwestern side of the galaxy \citep{Floyd2006,Labiano2014,RARiffel2023b}; together, these observations indicate that the northwest is the near side of the galaxy's disk, and the southeast is the far side. The residual velocity map for the molecular gas (third panel of Fig.~\ref{fig:rot_h2}) shows redshift excesses on the near side and blueshift excesses on the far side of the disk, associated with dust structures. Assuming the gas lies in the plane of the disk, these residuals can be interpreted as streaming motions toward the nucleus of 3C\:293. A schematic representation of the inner region of 3C\,293 is shown in Fig.~\ref{fig:schematic}. The diagram illustrates the orientation of the disk along with the proposed inflows (blue arrows), outflows (red arrows), and radio jet (green).
 
 Similar inflows of molecular and low-ionization gas associated with dust spiral arms on scales of hundreds of parsecs have been reported in other nearby galaxies \citep[e.g.,][]{Fathi06,Storchi-Bergmann07,RARiffel2008, RARiffel2013, Schnorr-Muller14, Davies2009, Davies2014}. Dust may trace shocks in the gas that cause a loss of angular momentum and consequent inflow toward the nucleus, feeding the AGN \citep{Brum2017}, or it may facilitate or enhance this inflow process, indicating that nuclear spiral arms represent possible fueling channels for the central SMBH and  AGN triggering \citep{Davies2014}. 

To estimate the inflow rate, we follow the approach of \citet{RARiffel2008}, assuming that the inflowing gas passes through a circular cross-section of radius $r$. The inflow rate is then given by 
\begin{equation}
    \dot{M}_{in} = 2 m_p N_{\rm H_2} v \pi r^2 n_{\rm arms}
\end{equation} where $m_p$ is the proton mass, $N_{\rm H2}$ is the molecular hydrogen density, and $v = v_{\rm obs}/\sin{\theta}$ is the inflow velocity in the plane of the galaxy. Here, $v_{\rm obs}$ is the observed velocity, and $\theta$ is the inclination angle of the disk, as determined from the rotation disk model. The parameter $n_{\rm arms}$ represents the number of spiral arms contributing to the inflow. In our case, we use $n_{\rm arms}$ = 2, representing the two high velocity residuals indicating by the arrows in Fig.~\ref{fig:rot_h2}. The cross-section radius is assumed to be half the width of the spiral arms, i.e, $r\approx0.3$ arcsec.

To estimate the density $N_{\rm H_2}$, we first derive the $\rm H_2$ gas mass of the disk. This requires knowing the gas temperature, which we calculate by \citep[e.g.][]{Wilman2005}:

\begin{equation}\label{temperature}
    \log \left( \frac{F_i \lambda_i}{A_i g_i}\right) = \mathrm{constant} - \frac{T_i}{T_{\rm exc}},
\end{equation} where $F_i$ is the extinction corrected flux of the $i-$th $\rm H_2$ transition, $\lambda_i$ is its corresponding wavelength, $A_i$ is the spontaneous emission coefficient, $g_i$ is the statistical weight of the upper level, $T_i$ is the upper level energy expressed as a temperature, and $T_{\rm exc}$ is the kinetic temperature under the assumption that the $\rm H_2$ gas is in local thermodynamic equilibrium. We measure the fluxes of the emission lines using a spectrum extracted within a circular aperture of 0.8 arcsec in diameter, by fitting each emission line by two gaussian components. By using the fluxes of the narrow component of all detected emission lines (18 in total) and fitting Eq.~\ref{temperature} to the data, we derive a temperature of $T_{\rm disk} = 2850 \rm \, K$ for the disk component

The hot $\rm H_2$ gas mass in the disk is subsequently determined using

\begin{equation}\label{H2_mass}
M_{\rm H_2} = \frac{2 m_{\rm p} F_{\rm H_2 2.12} 4 \pi D_{\rm L}^2 }{f_{\rm \nu = 1, J =3} A_{\rm S(1)} h \nu},
\end{equation} where $m_p$ is the proton mass, $F_{\rm H_2 2.12}$ is the extinction corrected flux of the $\rm H_2 \, 2.12\,  \mu m$ line, $ D_{\rm L}$ is the luminosity distance, $f_{\rm \nu = 1, J =3}$ is the upper-state population fraction, and  $A_{\rm S(1)}$ is the transition probability. For the derived disk temperature, the corresponding population fraction is $2.5 \times 10^{-2}$.

Then, we estimate $N_{\rm H_2}$ by
\begin{equation}
    N_{\rm H2} = \frac{M_{\rm H2}}{2m_p \pi r_{d}^2h},
\end{equation}
where $h$ is the height of the disk and $r_d$ is the radius of the disk. 

Adopting $r_d\approx$ 1.5 arcsec $\approx$ 1.5 kpc, a region that encompasses most of the H$_2$ emission, we obtain  $F_{\rm H_2 2.12}=1.09 \times 10^{-15} \, \rm erg\,s^{-1}$, resulting in $M_{\rm H_2} = 1.08 \times 10^{3} \, {\rm M_\odot}$. Then, using the inclination angle $\theta = 53^{\circ}$ determined by our rotational model (Sec.~\ref{subsec:diskprop}), we convert our observed velocity $v_{\rm obs} \approx $ 100\:km\:s$^{-1}$ (from Fig.~\ref{fig:rot_h2}) to an intrinsic inflow velocity $v = 100/\sin(53^\circ)\approx125$\:km\:s$^{-1}$. Combining this with the typical molecular disk height for active galaxies $h = 30$ pc \citep[e.g.][]{Hicks2009}, the estimated mass-inflow rate is $\dot{M}_{in} \approx 4 \times 10^{-4} \rm ~M_{\odot}\,yr^{-1}$.

This value is consistent with values reported for hot molecular gas in other nearby galaxies \citep[e.g.,][]{RARiffel2008,RARiffel2013,Diniz15}. However, it remains significantly lower than the values observed for low-ionization and cold molecular gas, which range from $10^{-2}$ to $10^{1} \, \rm M_{\odot} \, yr^{-1}$ \citep[e.g.][]{Storchi-Bergmann07,Muller-Sanchez09,Combes14,Audibert19,Audibert21,Roier22}. This aligns with the understanding that hot molecular gas constitutes merely a thin, heated layer of the much larger cold gas reservoir at the centers of galaxies \citep{dale05,ms06,mazzalay13}.

To assess whether this inflow is sufficient to power the AGN, we estimate the SMBH mass accretion rate using $\dot{m}_{\rm acc} = L_{bol}/(\eta c^2)$, where $L_{bol}$ is the bolometric luminosity, $c$ is the speed of light, and $\eta$ is the radiative efficiency of the accretion process. Adopting a bolometric luminosity range of $L_{\rm bol} = (2.5 - 7.5) \times 10^{43} \rm \,erg ~s^{-1}$ \citep{RARiffel2023b}, and a typical efficiency of $\eta = 0.1$, we obtain an accretion rate range of $\dot{m}_{\rm acc} = (4.4 - 12.6) \times 10^{-3} \:{\rm M_{\odot}\:yr^{-1}}$. This indicates that the inflow of hot molecular gas alone is not enough to fuel the SMBH in 3C\:293.

\subsection{Outflow properties}\label{subsec:outflowprop}

\begin{figure}
    \includegraphics[width=\linewidth]{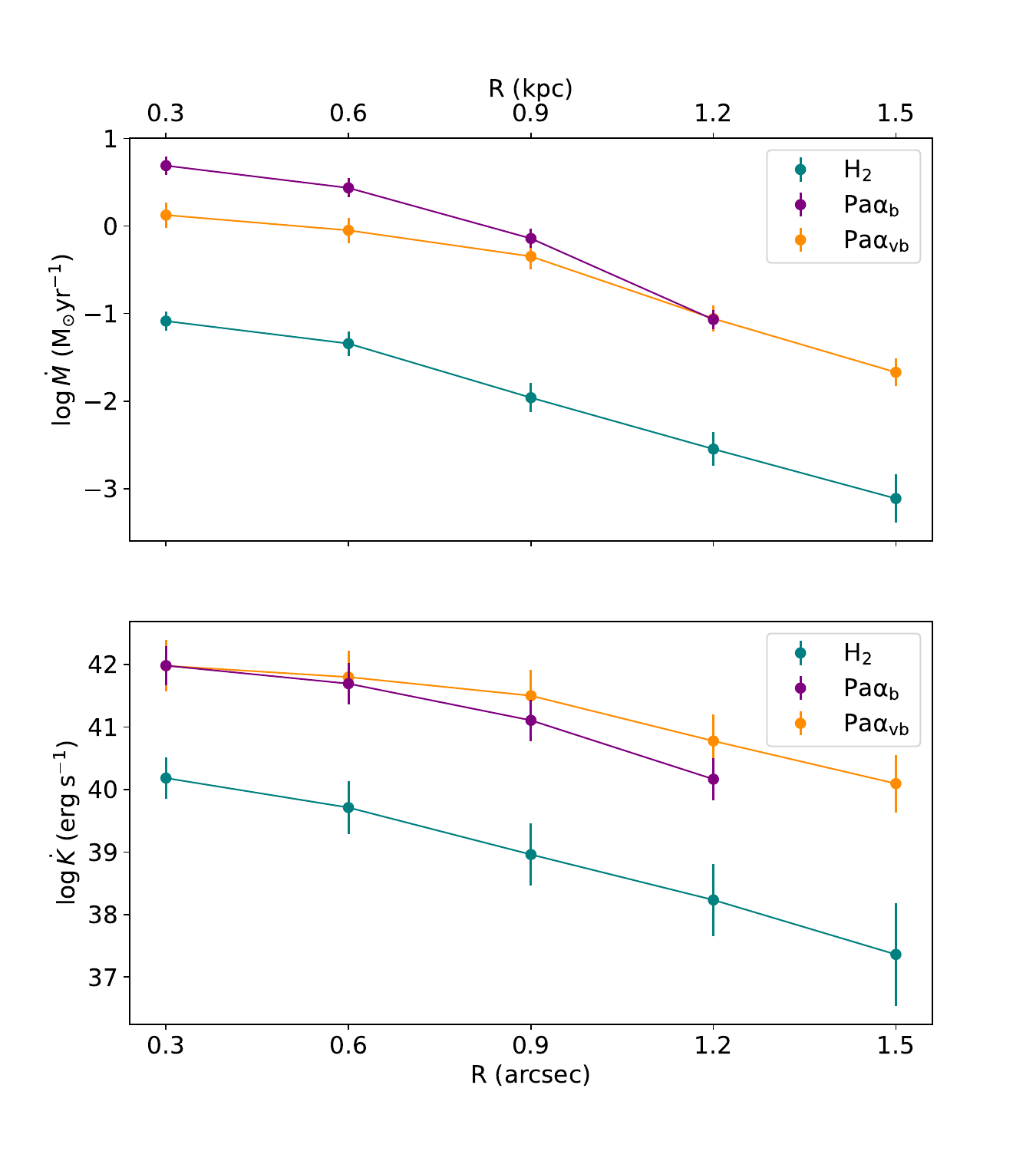}
    \caption{The top panel displays the radial distribution of the mass outflow rate, measured within radial bins of 0.3 arcsec width, shown on a logarithmic scale. Teal dots represent the hot molecular outflow, orange dots correspond to the outflow traced by the broad component of the Pa$\alpha$ emission line, and purple dots represent the fast outflow from the very broad component. The bottom panel presents the corresponding radial distributions of the kinetic power of the outflows.}
    \label{fig:mrate_kpower}
\end{figure}

We estimate the masses of hot molecular and ionized gas associated with the outflow from the observed fluxes of the $\rm H_2~2.12 \, \mu m $ and $\rm Pa\alpha$ emission lines, using only the flux from the broad components identified in our spectral fitting. To obtain a reliable estimate of the outflow properties, we excluded the central 0.1 arcsec region around the nucleus, due to its unresolved nature and the consequent lack of constraints on the radial gas distribution.

The derivation proceeds in two steps. First, we determine the temperature of the outflowing molecular gas using the same methodology as for the disk (Eq.~\ref{temperature}), applied to the broad line fluxes. This yields a temperature of $T_{\rm ouflow} = 3250\, \rm K$. Then, we calculate the gas mass using Eq.~\ref{H2_mass}, with a population fraction of $2.6 \times 10^{-2}$ for this temperature. 

For the ionized gas, on the other hand, the mass ($M_{\mathrm{H\,II}}$) is given by the equation

\begin{equation}
 \left( \frac{M_{\mathrm{H\,II} }}{M_{\odot}} \right) = 2.3 \times 10^{18} \left(\frac{F_{\rm Pa\alpha}}{\rm erg ~s^{-1} cm^{-2} } \right) \left(\frac{D}{\rm Mpc} \right)^2 \left(\frac{N_e}{\rm cm^{-3}} \right)^{-1} 
\end{equation}

\noindent where $F_{\rm Pa\alpha}$ is the extinction-corrected flux of the $\rm Pa\alpha$ emission line, and $N_e$ is the electron density, for which we adopt a value of 1000 $\rm cm^{-3}$, following previous studies \citep[e.g][]{RARiffel2023b, CostaSouza2024}.We derive a hot molecular gas outflow mass of $M_{\rm H_2} = \rm (6.0 \pm 0.016) \times 10^{4} \,M_{\odot}$, while for the ionized gas, we find masses of $M_{\mathrm{H\,II}, \rm b} = \rm (3.0 \, \pm \, 0.02) \times 10^6 \, M_{\odot}$ and $M_{\mathrm{H\,II}, \rm vb} = \rm (5.8 \, \pm \, 0.5) \times 10^5 \, M_{\odot}$ for the broad and very broad components, respectively. 

We estimate the mass-outflow rate ($\dot{M}$) and the kinematic power ($\dot{K}$) of the molecular and ionized outflow assuming a shell-like geometry, with width $\Delta R = 0.3$ arcsec, in alignment with the methodology adopted in \citet{RARiffel2023b} and \citet{CostaSouza2024} for the ionized and warm molecular gas.  The $\dot{M}$ and $\dot{K}$ are obtained using the following expressions

\begin{align}
    &\dot{M} = \frac{M_{\rm out, \Delta R} V_{\rm out, \Delta R}}{\Delta R},  \\
    &\dot{K} = \frac{M_{\rm out, \Delta R} (V_{\rm out, \Delta R})^3}{2\Delta R},
\end{align} where $M_{\rm out, \Delta R}$ is the mass of the outflowing gas within each shell, and $V_{\rm out, \Delta R}$ is the outflow velocity, defined by

\begin{equation}
    V_{\rm out, \Delta R} = \lvert \langle V_{\rm broad} \rangle _{\Delta R}\rvert + 2 \langle \sigma_{\rm broad} \rangle _{\Delta R},
\end{equation} with $\langle V_{\rm broad}\rangle _{\Delta R}$ representing the mean velocity of the outflow in each shell, and $\langle \sigma_{\rm broad} \rangle _{\Delta R}$ its mean velocity dispersion within each shell. 

We applied this procedure to calculate the $\dot{M}$ and $\dot{K}$ for both the molecular and ionized outflows. Fig.~\ref{fig:mrate_kpower} presents the results of each shell for the mass-outflow rate and the kinetic power, in the top and in the bottom panel, respectively.  The highest values of $\dot{M}$ and $\dot{K}$ are observed in the ionized gas, traced by the broad component of the Pa$\alpha$ emission line. The mass-outflow rate reaches a peak of $4.9 \pm 1.2 ~\rm M_{\odot}~yr^{-1}$, while the kinetic power peaks at $9.6 \times 10^{41} \rm \,erg~s^{-1}$. In comparison, the very broad $\rm Pa \alpha$ (fast outflow) component shows peak values of $ 1.6 \pm 0.5 ~\rm M_{\odot}~yr^{-1}$ for the mass-outflow rate and $1.14 \times 10^{42} \rm \,erg~s^{-1}$ for the kinetic power. For the molecular outflow, the maximum mass-outflow rate is $ 0.08 \pm 0.02 ~\rm M_{\odot}~yr^{-1}$, with a corresponding kinetic power of $1.53 \times 10^{40} \rm \,erg~s^{-1}$.

The hot molecular and ionized gas mass-outflow rates can be compared with previous estimates for 3C\:293. \citet{CostaSouza2024}, using JWST/MIRI observations, estimated a mass outflow rate of 4.90$\, \pm\, $2.04 M$_\odot$\:yr$^{-1}$ for the warm (198 K $\leq T\leq$ 1000 K) molecular gas, and 1.22$\,\pm\,$0.21 M$_\odot$\:yr$^{-1}$ for the hot (1000 K $\leq T\leq$ 5000 K) molecular gas, while \citet{RARiffel2023b} reported an ionized gas mass outflow rate of 0.5$\,\pm\,$0.1 M$_\odot$\:yr$^{-1}$. The mass outflow rate obtained for the ionized gas here is about one order of magnitude higher than that reported by \citet{RARiffel2023b} using optical observations, which may be attributed to the lower extinction in the near infrared, allowing the detection of outflows in denser regions obscured by dust. On the other hand, the hot molecular outflow rate estimated here is about one order of magnitude lower than the value obtained by \citet{CostaSouza2024} using mid-infrared H$_2$ emission lines. This difference could also be related to the lower interstellar extinction in the mid-infrared compared to the near-infrared. However, it may also arise from the contribution of lower-temperature gas, as \citet{CostaSouza2024} integrated the outflow mass for temperatures starting at 1000 K, whereas we used a temperature of $\sim \,$ 3250 K to estimate the mass.

The outflows from an AGN can be driven either by radiation fields -- referred to as radiative or wind mode -- or by relativistic jets -- what is known as kinematic or jet mode feedback \citep{Fabian2012}. In the radiative mode, the matter accreted onto the SMBH converts gravitational potential energy into electromagnetic radiation and kinetic energy, in part through the ejection of particles in the form of winds. These radiatively driven winds primarily couple to the hot, diffuse ISM, requiring only about 0.5\% of the AGNs bolometric luminosity to regulate star formation \citep{Hopkins2010}. This initial coupling allows the outflow to interact with cold molecular clouds, being capable of suppress star formation on galactic scales. In the kinetic mode, on the other hand, relativistic jets launched from the AGN propagate through the ISM and couple with the dense gas. Numerical simulations suggest that up to 20\% of the jet's kinetic power can be transferred to the surrounding medium \citep{Wagner2012, Mukherjee2016}. On small scales ($\lesssim \, 1 \rm \: kpc$), the interaction between the jet and the ISM can produce ram-pressure shocks and even form cavities \citep{Dugan2017, Mandal2021}. Observational studies have widely reported jet-ISM interactions as key mechanisms behind AGN-driven feedback \citep[e.g][]{Morganti2015, Nesvadba2017, Audibert2023, Duncan2023}. Simulations exploring radiative, kinetic, and hybrid AGN feedback suggest that the presence of jets plays a dominant role in the suppression of star formation, with jet-coupling efficiencies as low as 1\% being sufficient to suppress stellar activity in the host galaxy \citep{Husko2024}.

Assuming a radiatively driven outflow, we estimated the maximum kinetic coupling efficiency ($\epsilon$) as the ratio between the maximum $\dot{K}$ and the AGN bolometric luminosity, given by $L_{\rm bol} = (2.5 - 7.5) \times 10^{43} \rm erg ~s^{-1}$ \citep{RARiffel2023b}. For the ionized gas, traced by the broad and very broad components of the $\rm Pa\alpha$ emission line we obtained $\epsilon$ values of $(1.3 - 3.8)\%$ and $(1.6 - 4.6)\%$, respectively, consistent with previous estimates by \citet{RARiffel2023b} and \citet{CostaSouza2024}. These estimates exceed the minimum thresholds suggested by cosmological simulations for AGN feedback to effectively transfer energy to the ISM and regulate star formation \citep[e.g][]{Hopkins2010,Harrison2018}.

Although these results indicate that radiative feedback is energetically capable of affecting the ISM, several studies point to the radio jet as the dominant mechanism driving the outflow in the 3C\,293 \citep{mahony13,Lanz2015,Mahony2016,RARiffel2023b,CostaSouza2024,RARiffel2025}. To quantify the efficiency of coupling of the jet-driven outflow, we calculate the ratio between the maximum $\dot{K}$ of the outflow and the jet power, which is estimated to be $(2 – 4) \times 10^{43}\, \rm erg \, s^{-1}$ based on the 151 MHz radio luminosity and the spatial extent of the radio lobes \citep{Lanz2015}. Considering both broad components of the $\rm Pa\alpha$ emission line, we find coupling efficiencies of (2.4 - 4.8)\% for the broad component, and (2.85 - 5.7)\% for the very broad one. These values are within the range predicted by jet-ISM interaction models \citep[e.g][]{Wagner2011, Cielo2014, Talbot2022, Husko2024},  indicating that the jet-driven outflows are powerful enough to affect the star formation in the host galaxy.

\section{Conclusions}\label{sec:conclusions}

We used integral field spectroscopy with the JWST NIRSpec instrument, covering the spectral range from $\sim1.6$ to $\sim3.0\:\mu$m, to map the stellar kinematics, emission structure and kinematics of hot molecular (using the H$_2$\:2.12$\,\mu$m emission line) and ionized gas (traced by the [Fe\,{\sc ii}]\:1.64$\,\mu$m and Pa$\alpha$ emission lines), and gas extinction in the central $\sim3\times3\,{\rm kpc^2}$ of the radio galaxy 3C\:293, with a spatial resolution of $\sim 130$ pc.  Our main findings are as follows:

\begin{itemize}
    \item The stellar velocity field exhibits a well defined rotation pattern, with the line of nodes aligned at $\psi_0 = 46^{\circ}\pm7^{\circ}$, consistent with the orientation of the large scale disk. The kinematic center is shifted by about $\sim0.5$~arcsec from the near-infrared emission peak, potentially as a result of a recent merger event. The stellar velocity dispersion map predominantly shows low values ($\sigma \lesssim 100~{\rm km\,s^{-1}}$) in the eastern region, likely associated with an intermediate-age stellar population, while higher values (up to  $180~{\rm km\,s^{-1}}$) are observed at the nucleus and to the south-west, attributed to bulge stars.

    \item The ionized gas exhibits three main kinematic components: one associated with disk emission and two produced by outflows, identified as narrow ($\sigma \sim 100 \rm \, km \, s^{-1}$ ), broad ($\sigma \sim 250  \rm \, km \, s^{-1}$), and very broad ($\sigma \sim 640 \rm \, km \, s^{-1}$) emission line components. For the hot molecular gas, only two components are observed: one resulting from the motions of the gas in the disk and another associated with outflows. 

    \item Similarly to the stars, the H$_2$ velocity field is overall well reproduced by a rotation disk model. However, the analysis of the residual map reveals systematic redshifted residuals on the northwest side of the galaxy and blueshifted residuals on the southeast side, corresponding to the near and far sides of the disk, respectively. These residuals are co-spatial with dust lanes observed in HST images and are interpreted as inflows toward the center. We estimate a mass-inflow rate in hot molecular gas of $\dot{M}_{in} \approx 4 \times 10^{-4} \rm ~M_{\odot} yr^{-1}$, which is not enough to sustain the AGN of 3C\,293 at its current luminosity. 

    \item Using the observed Pa$\alpha$ emission and kinematics for the outflowing components, we estimate an ionized gas mass outflow rate of $4.9 \pm 1.2 ~\rm M_{\odot}~yr^{-1}$ using the broad component, and $ 1.6 \pm 0.5 ~\rm M_{\odot}~yr^{-1}$ for the very broad component. The outflow rate in hot molecular gas is much lower, of $ 0.08 \pm 0.02 ~\rm M_{\odot}~yr^{-1}$, consistent with the fact that the hot molecular gas represents only a hot skin of the total molecular gas in the central region of galaxies. 

    \item The outflows in 3C\,293 are likely driven by the interaction between the radio jet and the interstellar medium, as they are co-spatial with the radio core and exhibit kinetic powers approximately one order of magnitude lower than the jet power. Although our calculations show that a radiatively driven scenario is also plausible, with a kinetic coupling efficiency of 1-4\%, the spatial correlation point the radio jet as the dominant mechanism. The maximum kinetic coupling efficiency of the jet-driven ionized outflow is around 5.7\%, which exceeds the thresholds predicted by simulations for AGN feedback to effectively suppress star formation.

    \item The gas emission is highly attenuated by dust, as indicated by the observed Br$\beta$/Pa$\alpha$ line ratio. The highest extinction, with values of up to $A_V\approx35$, is observed for the outflow component. The extinction map for the disk component present the highest values in regions co-spatial with the strong dust lanes seen in HST images.  This high dust attenuation can explain the offset between the optical and infrared nucleus of 3C\,293.

\end{itemize}

In summary, our results provide new evidence of multi-phase outflows from the nucleus of 3C\,293. In addition, we identified streaming motions towards the nucleus, which may result in a gas reservoir to feed the central AGN. This work highlights the power of JWST/NIRSpec high-sensitivity observations to study highly obscured systems, providing access to the dustier and denser gas phases, which are essential for improving our understanding of AGN feeding and feedback processes. 
In future work, we will focus on investigating the origin of $\rm H_2$ and ionized gas emission in 3C\,293, as well as their relationship with the radio jet and outflows.

%\section{Software and third party data repository citations} %\label{sec:cite}

%% Also note that the akcnowlodgment environment does not support long amounts of text. If you have a lot of people and institutions to acknowledge, do not use this command. Instead, create a new \section{Acknowledgments}.

%\begin{acknowledgments}
\section*{Acknowledgments}
We thank the referee for their valuable comments that helped us to improve our manuscript.
MSZM acknowledges financial support from Coordena\c c\~ao de Aperfei\c coamento de Pessoal de N\'ivel Superior (CAPES; Finance Code 001). 
RAR acknowledges the support from Conselho Nacional de Desenvolvimento Cient\'ifico e Tecnol\'ogico (CNPq; Proj. 303450/2022-3, 403398/2023-1, \& 441722/2023-7) and CAPES (Proj. 88887.894973/2023-00). 
GLSO and HCPS thank the financial support from CAPES (Finance Code 001) and CNPq. 
N.L.Z. is supported in part by NASA through STScI grant JWST-ERS-01928.
MB thanks the financial support from the IAU-Gruber foundation fellowship.
TSB acknowledges financial support from CNPq (Projs. 425966/2016-0 and 303450/2022-3) and Funda\c{c}\~ao de amparo \`{a} pesquisa do Rio Grande do Sul (FAPERGS; Projs. 16/2551-0000495-1 and 21/2551-0002018-0).
RR acknowledges support from CNPq (Proj. CNPq-445231/2024-6,311223/2020-6, 404238/2021-1, and 310413/2025-7), FAPERGS (Proj. 19/1750-2 and 24/2551-0001282-6) and CAPES (Proj. 88881.109987/2025-01).

 %RR acknowledges support from  Conselho Nacional de Desenvolvimento Cient\'{i}fico e Tecnol\'ogico  ( CNPq, Proj. CNPq-445231/2024-6,311223/2020-6, 404238/2021-1, and 310413/2025-7), Funda\c{c}\~ao de amparo \`{a} pesquisa do Rio Grande do Sul (FAPERGS, Proj. 19/1750-2 and 24/2551-0001282-6) and Coordena\c{c}\~ao de Aperfei\c{c}oamento de Pessoal de N\'{i}vel Superior (CAPES, 88881.109987/2025-01).

%\end{acknowledgments}


\section*{Data Availability}
The data utilized in this study originate from the JWST Cycle 1 program (ID 1928). The full dataset is publicly available through the Mikulski Archive for Space Telescopes (MAST) hosted by the Space Telescope Science Institute. It can be accessed via \dataset[DOI: 10.17909/tazj-hp44]{https://doi.org/10.17909/tazj-hp44}.

\vspace{5mm}
\facilities{JWST(NIRSPec)}


\software{astropy \citep{2013A&A...558A..33A,2018AJ....156..123A},
IFSCUBE  \citep{RuschelDutra2021}
pPXF \citep{Cappellari2004, Cappellari2017, Cappellari2023}
%          Cloudy \citep{2013RMxAA..49..137F}, 
%          Source Extractor \citep{1996A&AS..117..393B}
          }




\bibliography{paper}{}
\bibliographystyle{aasjournal}



\end{document}